\documentclass[fleqn,usenatbib]{mnras}

\usepackage[T1]{fontenc}

\DeclareRobustCommand{\VAN}[3]{#2}
\let\VANthebibliography\thebibliography
\def\thebibliography{\DeclareRobustCommand{\VAN}[3]{##3}\VANthebibliography}

\usepackage{graphicx}	
\usepackage{amsmath}	
\usepackage{amssymb}	
\usepackage{subcaption}

\usepackage{xcolor}
\definecolor{arpcolor}{RGB}{51, 166, 255}

\usepackage{newtxtext,newtxmath}

\title[bar origins, structure \& migration]{The impact of bar origin and morphology on stellar migration }

\author[E. J. Iles et al.]{
Elizabeth J. Iles,$^{1,2}$\thanks{E-mail: elizabeth.iles@sydney.edu.au}
Alex R. Pettitt,$^{3}$
Takashi Okamoto$^{1}$
and Daisuke Kawata $^{4}$
\\
$^{1}$Department of Physics, Faculty of Science, Hokkaido University, Sapporo 060-0810, Japan\\
$^{2}$Sydney Institute for Astronomy, School of Physics, University of Sydney, Camperdown, NSW, 2006, Australia\\
$^{3}$Department of Physics and Astronomy, California State University, Sacramento, 6000 J Street, Sacramento, CA 95819-6041, USA\\
$^{4}$Department of Space \& Climate Physics, University College London, 124 Holmbury St Mary, Dorking, RH5 6NT, UK\\
}

\date{Accepted XXX. Received YYY, in original form ZZZ}

\pubyear{2023}

\begin{document}
\label{firstpage}
\pagerange{\pageref{firstpage}--\pageref{lastpage}}
\maketitle

\begin{abstract}
Different mechanisms driving bar structure formation indicate that bar origins should be distinguishable in the stellar populations of galaxies. To study how these origins affect different bar morphologies and impact stellar orbits and migration, we analyse three simulated discs which are representative of bar formation under isolated evolution motivated by disc instability, and interaction driven tidal development. The first isolated disc and the tidally driven disc produce similar bar structure, while the second isolated disc, generated by the tidal initial condition without the companion, is visibly dissimilar. Changes to radial and vertical positions, angular momentum in the disc-plane, orbital eccentricity and the subsequent disc metallicities are assessed, as is the dependence on stellar age and formation radii. Bar origin is distinguishable, with the tidal disc displaying larger migration overall, higher metallicity difference between the inner and outer disc, as well as a population of inner disc stars displaced to large radii and below the disc-plane. The affect of closest approach on populations of stars formed before, after and during this period is evident. However, bar morphology is also found to be a significant factor in the evolution of disc stellar properties, with similar bars producing similar traits in migration tendency with radius, particularly in vertical stellar motion and in the evolution of central metallicity features. 

\end{abstract}


\section{Introduction}
\label{s:intro}
Radial migration is the term often used to describe the processes capable of displacing stars over large radial distances ($\sim$kpc) and restructuring a galaxy over time \citep[e.g.][]{Sellwood2002}. It was initially identified through angular momentum changes at co-rotation between the pattern speeds of stars and spiral arms, as well as at Lindblad resonances \citep{Lynden-Bell1972}. When discussing radial migration, many refer to processes of \textquoteleft blurring\textquoteright \ and \textquoteleft churning\textquoteright \ to describe the two possible mechanisms for stars born in other regions of the disc to travel to and contaminate the local properties in a given other region \citep[e.g.][]{Sellwood2002, Schonrich2009b}. Regardless of whether this change is permanent, and thus considered migration, blurring refers to the case where the amplitude of radial oscillations around an average guiding radius ($R_g$) for the orbit changes. If any given star in the galaxy is considered to begin on a roughly circular orbit from which it is later scattered, such that the orbit becomes radially extended but the guiding radius and thus, the angular momentum ($L_z$) is unaltered, we call this blurring. Comparatively, churning is a process triggered by torques arising from non-axisymmetric features in the disc, such as bars and spiral arms, causing changes in the angular momenta of stars on a given orbit but with no associated change to the orbital eccentricity \citep{Sellwood2002, Schonrich2009b}. In general, both processes are responsible for transporting stars radially across the disc \citep[e.g.][]{Schonrich2009a, Loebman2011, Roskar2012, Buck2020}. These displaced stars then serve to mix chemically distinct components which would otherwise follow a single evolution track in a given region and thus, contribute to the complex chemical features observable in the Milky Way, as well as many similar external galaxies \citep[e.g.][]{Roskar2008, Schonrich2009a, Minchev2013}.

However, despite being a relatively well-determined process, it is still difficult to comprehensively determine the conditions within a disc which may promote a given migration. Which stars in the galaxy are likely to undergo migration and which of the two process will dominate, remains not well-understood, although many have attempted to quantify this via analytical methods \citep[e.g.][]{Sellwood2002, Schonrich2009a, Schonrich2009b, Schonrich2017}, numerical simulations \citep[e.g.][]{Roskar2008, Quillen2009, Halle2015, Aumer2017a, Mikkola2020} and, of course, various observational surveys \citep[e.g.][]{Minchev2018, Frankel2018}. These studies have been predominantly concerned with determining the extent of the migration, usually as a function of stellar position and/or velocity relative to a mid-plane \citep[e.g.][]{Schonrich2009a,Schonrich2009b, Solway2012a, Vera-Ciro2014, Vera-Ciro2016a}, while other metrics have been considered, such as dynamical temperature \citep[e.g.][]{Daniel2018} or vertical/radial action \citep[e.g.][]{Mikkola2020}. The relative effects of varying disc structures, such as barred and non-barred spirals, have also been a major focus \citep[e.g.][]{Silchenko2010, Grand2012, DiMatteo2013, Kawata2017, Lin2017, Halle2018}, with bars being of keen interest due to the strong torques they induce which may trigger such radial migration. Radial migration studies directed at interacting discs \citep[e.g.][]{Quillen2009, Zinchenko2015, Buck2020}, with the added complexity of tidal effects, are relatively rare but these are also necessary for a comprehensive understanding of real galaxies, as it is unavoidable that in the $\Lambda_{\rm CDM}$ universe structure forms hierarchically \citep{Abadi2003, Hopkins2010}. It is even less common, however, to find studies which consider the evolutionary histories of these galaxies and consequently, whether the processes driving the formation of non-axisymmetric disc structures may also cause the stellar motions and migration to differ. 

In terms of significant observable effects, radial migration has been previously associated with a range of key features including, but not limited to: flat age-metallicity relations \citep[e.g.][]{Casagrande2016}; the metallicity-rotation velocity relation \citep[e.g.][]{AllendePrieto2016, Kordopatis2017, Schonrich2017}; mono-age population flaring in the outer-disc \citep[e.g.][]{Minchev2012b, Minchev2014}; and as a means for driving the [$\alpha$/Fe]$-$[Fe/H] bi-modality of stars into spatially distinct thick and thin disc structural components \citep[e.g.][]{Schonrich2009a}. Although, the results of many of these studies show migration alone--or even at all--may not be sufficient to explain these features, significant work is still ongoing. This is particularly true in galaxies with complex evolutionary and interaction-driven histories, as interactions often affect similar disc properties. For example, both migration and mergers have previously been considered to affect population flaring, although the impact of migration was not necessarily found to occur in the same regions, to the same extent--or even at all, in some cases--compared to interaction-related flaring \citep[e.g.][]{Villalobos2008, Bournaud2009, Minchev2015}. This supports the imperative to determine whether migration is affected differently in discs with a range of evolutionary histories, as many of the significant features observed in galaxies are more likely to be caused by the intersection of many factors rather than just one singular process.

Within the many variations of galactic morphology observable in the universe, the barred-spiral features in a significant fraction of galaxy discs $\sim$25-75\%, depending on the classification criteria \citep[e.g.][]{Schinnerer2002, Aguerri2009, Masters2011}. The Milky Way is also expected to host to one such feature \citep[e.g.][]{Blitz1991, Nakada1991, Paczynski1994, Zhao1994}, which makes understanding the formation and evolution of these features, as well as their impact on the host-galaxy, particularly important. Previous studies have long demonstrated that these bar-like features can form in the central regions of a spiral for kinematically cold, sufficiently massive stellar discs, evolving independently in relative isolation \citep[e,g.][]{Hohl1971, Ostriker1973} or alternatively, be induced into forming and altering the disc structure due to the influence of external tidal forces, via both major and minor galaxy-galaxy interactions \citep[e.g.][]{Noguchi1987, Salo1991}. In the isolated evolutionary scenario, an initial instability in the disc will trigger a phase of rapid development in which the bar emerges, subsequently buckling in and out of the disc-plane for a brief period, before slowing and growing gradually over a more prolonged phase of secular evolution \citep[e.g.][]{Raha1991, Sellwood2014}. Comparatively, the effects of interactions are more varied. It has been shown that interactions are able to induce bar formation in isolated discs which should have otherwise been stable against bar-forming instabilities, as well as to effect bar formation in discs which were already able to independently develop bars \citep[e.g.][]{Romano-Diaz2008, Mendez-Abreu2012, Lang2014, Moetazedian2017}. Certain interaction conditions can also dampen or completely halt bar formation, rather than induce or drive it \citep{Athanassoula2002, Kyziropoulos2016, Gajda2017, Moetazedian2017, Zana2019}. It has additionally been shown that bars driven by interactions should rotate slower than those formed in isolation \citep{Miwa1998, Martinez-Valpuesta2017, lokas2018}. There is currently no consensus on which scenario is most likely, or even whether naturally occurring bars may, in fact, form under both possible mechanisms depending on the environment of the host-galaxy \citep{Skibba2012, Cavanagh2022}. 

While the debate on the origins of these features continues, many studies are also focused on the relative impact of galactic bars on observable properties of the host-galaxy \citep[e.g.][]{Roberts1979, Athanassoula1992, Downes1996, Sheth2002, Emsellem2014, Beuther2018, Watanabe2019}. However, while these studies increasingly demonstrate how the presence of a bar will affect various observable properties, this is often considered independently of the mechanisms contributing to bar formation within a given galaxy. The following results are part of an endeavour to determine whether the different mechanisms capable of producing barred-spiral structure in galaxies, may also subsequently effect the impact of these structures on the host-stellar populations in discernibly different ways. We make use of Smoothed Particle Hydrodynamics (SPH) simulations with the \textsc{Gasoline2} code \citep{Wadsley2004, Wadsley2017} to produce a small sample of barred-galaxies to study. These are designed to be representative of the two possible bar formation mechanisms (isolated \& tidally-driven) and are initialised from measurements of nearby, resolved galaxies for observational consistency. The presentation of this work is as follows: an outline of the simulation specifics in Section \ref{s:ICs}; in Section \ref{s:results}, the primary results and analysis; and finally, the main conclusions in Section \ref{s:conc}.

\section{Simulation Parameters}
\label{s:ICs}
Three simulated discs are produced and analysed over the relatively early phase of bar evolution ($\le$1 Gyr) to consider spatially and temporally varying trends associated with developing barred-disc structures, as triggered by different mechanisms. These discs are IsoB, a barred-galaxy with an isolated evolutionary history tailored to measurements of NGC\,4303; the similarly barred TideB, which is tailored to NGC\,3627 and externally driven by the tidal forces of a smaller companion in a minor merger-like interaction; and TideNC, a third disc which serves as a comparison where the initial condition of TideB is evolved without the influence of the companion. The development of the first two discs (IsoB \& TideB), along with the initial conditions used and simulation parameter selection, is discussed in \citet{Iles2022}, focusing on the star formation properties between these discs. Consequently, the introduction here remains brief.

The simulated discs, each evolving into barred-spiral morphologies, were produced from initial conditions generated from the \textsc{GalIC} package \citep[IsoB;][]{Yurin2014} and previous interaction studies \citep[TideB, TideNC;][]{Pettitt2018} and constrained to align with surface density profiles and kinematic data of the target galaxies \citep[NGC\,4303 \& NGC\,3627;][]{Iles2022}. These discs can, therefore, be considered indicative of either isolated bar evolution, motivated by disc instability (IsoB, TideNC), or tidally-driven bar evolution, arising from an interaction triggered early bar development (TideB). All three discs have gas mass resolutions of approximately 1000\,M$_\odot$ (IsoB $\sim$1044\,M$_\odot$; TideB, TideNC $\sim$1084\,M$_\odot$) and active $N$-body particle components for each bulge and disc stars, gas and dark matter. However, the gas component in these discs should be considered singular and does not specifically differentiate between molecular and atomic states. The relative mass and scale length parameters, corresponding to exponential discs, Hernquist bulges, and NFW-like halos (see \citet{Yurin2014} for details), are listed in Table \ref{t:ic} for reference. 

The simulation parameters are held consistent for the evolution of these discs over a period of 1\,Gyr in the \textsc{gasoline2} SPH environment with the standard hydrodynamical treatment advocated by \citet{Wadsley2017} with 200 neighbours and a Wendland C4 kernel \citep{Dehnen2012}. Gravitational softening lengths are prescribed for each component with values of 0.1\,kpc for the halo, 0.05 kpc for stars and 0.01\,kpc for gas. A temperature threshold of 300\,K, density threshold of 100\,atoms/cc and a convergent flow requirement are the primary conditions for star formation with a star forming efficiency of 10\% (C$_\star$ = 0.1) and a \citet{Chabrier2003} IMF, consistent with standard sub-grid prescriptions \citep{Katz1996, Stinson2006, Wadsley2004, Wadsley2017}. The implementation of UV and photoelectric heating, as well as metal cooling in the form of a tabulated cooling function \citep{Shen2010}, recovers a two-phase thermal profile comparable to the ISM \citep{Wolfire2003} from an initially isothermal ($10^4$K) gas profile. Stellar feedback is implemented from supernova following the super-bubble method of \citet{Keller2014}.

\begin{table}
	\centering
	\caption{Mass (in units of $10^{10}$\,M$_\odot$) and scale length (in units of kpc) for the initial condition of each simulated disc galaxy, with the distance parameter for the companion defined by closest approach (in kpc).}
	\label{t:ic}
	\begin{tabular}{lccccccccc} 
		\hline
		&\,$M_{_{\rm gas}}$ &\,$M_{*_{\rm disc}}$ &\,$M_{*_{\rm bulge}}$ &\,$M_{_{\rm halo}}$ &\,$M_{_{\rm companion}}$ \\
		\hline
		\textbf{IsoB} & 0.522 & 2.611 & 0.402 & 37.16 & - \\
		\textbf{TideB} & 0.759 & 2.441 & 0.072 & 42.57 & 2.401\\
		\textbf{TideNC} & 0.759 & 2.441 & 0.072 & 42.57 & -\\
		\hline
		& $a_{_{\rm gas}}$ & $a_{*_{\rm disc}}$ & $a_{*_{\rm bulge}}$ & $a_{_{\rm halo}}$ & $b_{_{\rm companion}}$ \\
		\hline
		\textbf{IsoB} & 3.090 & 2.060 & 2.057 & 20.57 & -\\
		\textbf{TideB} & 3.705 & 2.470 & 0.405 & 20.26 & 10 \\
		\textbf{TideNC} & 3.705 & 2.470 & 0.405 & 20.26 & - \\
		\hline
	\end{tabular}
\end{table}

\begin{figure*}
  \centering
	\includegraphics[width=0.92\textwidth]{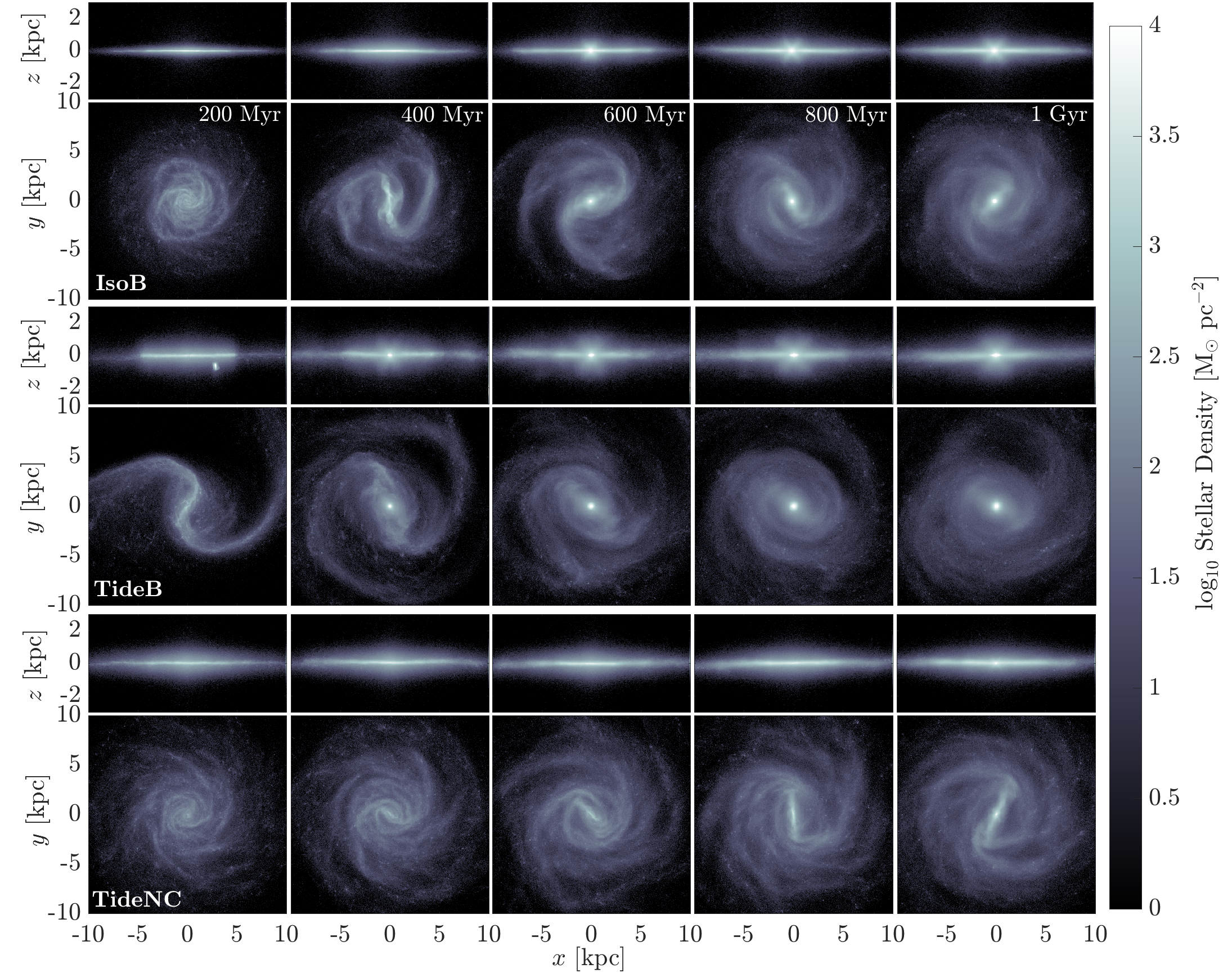}
  \caption{Projection of face-on and side-on stellar density distributions, set into the $xy$ and $xz$-plane respectively, for each: IsoB, TideB and TideNC. Columns step in time by 200\,Myr to the total simulation time of 1\,Gyr.}
  \label{f:starprj_xzy}
\end{figure*}

The period of 1\,Gyr is intentionally designed to capture the disc evolution, wherein the effects of the tidal perturbation are most prominent in the interacting disc, thus highlighting any fundamental differences arising between the isolated and tidally-driven bar origins. The evolution of the stellar component for each of these discs can be seen in Figure \ref{f:starprj_xzy}, both in face-on and side-on orientations. The colour weighting accounts for the stellar mass density while time progresses through each column in periods of 200\,Myr simulation time. To describe each disc in terms of general features, both the IsoB and TideB discs form similar bars and primarily two-arm spiral structures, although the formation timescales for these features is slightly different, as the interaction of the tidal disc drives bar formation $\sim$200\,Myr faster than the IsoB disc with the isolated environmental conditions \citep{Iles2022}. The bar lengths and strengths appear visually similar within these two discs. The number and prominence of the arms, as well as their pitch angles, are also similar. The third disc (TideNC) can also be seen to form an obvious bar feature during the later stages of the 1\,Gyr period, despite the relatively higher stability criterion in the Tide initial condition. The bar feature exhibited by this disc is, however, visibly different in morphology to either of the other bars (in IsoB or TideB). This is particularly evident in the side-on density distribution. In the central regions of both IsoB and TideB, the disc vertical structure exhibits an obvious cross- or X-shape symmetrically above and below the disc-plane during the barred periods. Such morphology is often associated with bar formation and the warping of the bar in and out of the disc-plane as it forms \citep{Raha1991,lokas2018, Sellwood2020}. While the face-on projections of TideNC also appear to exhibit clear bar-like morphology in the central regions, no similar feature is discernible in the side-on projections for this disc, though there is a faint X-shape feature building up in the last panel, and a X-shaped bulge may grow if a longer evolution is simulated. This directly implies a significant difference exists between the bars formed in TideNC and those of IsoB and TideB. It also affirms the interaction as the primary influence on bar formation in TideB, as intended.

\section{Results}
\label{s:results}
The following analysis is aimed at identifying trends in stellar dynamics characterised by changes in position, momenta and stellar orbits over 1\,Gyr of disc evolution. This will facilitate an investigation into whether these trends may be sufficiently sensitive to bar origin so as to impact lasting and observable features linking the effects of radial migration with the formation mechanisms and evolutionary histories of galaxies. 

\subsection{Standard Migration Signatures}

Radial migration is traditionally quantified by the changes in either or both radial position ($r$) and orbital angular momentum in the disc plane ($L_z$), which also allows for a distinction between the two processes of blurring and churning thought to drive such changes \citep[e.g.][]{Sellwood2002}. It has also been argued that vertical motion, which occurs perpendicular to the disc plane (in direction $z$), is not without influence in studies of stellar population mixing, particularly in relation to the formation of distinct thick and thin disc structures \citep[e.g.][]{Navarro2018, Mikkola2020}. Whether these simulations would be sufficiently able to reproduce and resolve this vertical structure was not considered at the time of determining the initial conditions for these simulated galaxies, as we were aiming to predominantly focus on features within the disc-plane. However, it has been possible to include the variation for all three of these key parameters in the following analysis to probe for potential signatures of bar origins in galaxies, even post-interaction. Additionally, we note that these results focus on tracing only stars formed within the simulation and exclude the less self-consistent initial stellar population. This is driven by the motivation to relate evolving disc structure with its associated observable stellar features. For clarity in the following discussion, we also define an \textquoteleft initial\textquoteright \ value to be the value of a parameter for a given star at the specific commencement time of the period of interest (or at star formation if stars formed during this period are included), while \textquoteleft final\textquoteright \ is the corresponding value for the same star at the end of the period of interest. To later differentiate this value from the initial values of these parameters at the time of formation for a given star, we will use the subscript \textquoteleft form\textquoteright \ to indicate that the parameter is measured specifically from the time of star formation (i.e. initial for the period: $t_i, r_i$ vs. star formation: $t_{\rm form}, r_{\rm form}$). 

\begin{figure*}
  \centering
	\includegraphics[width=0.82\textwidth]{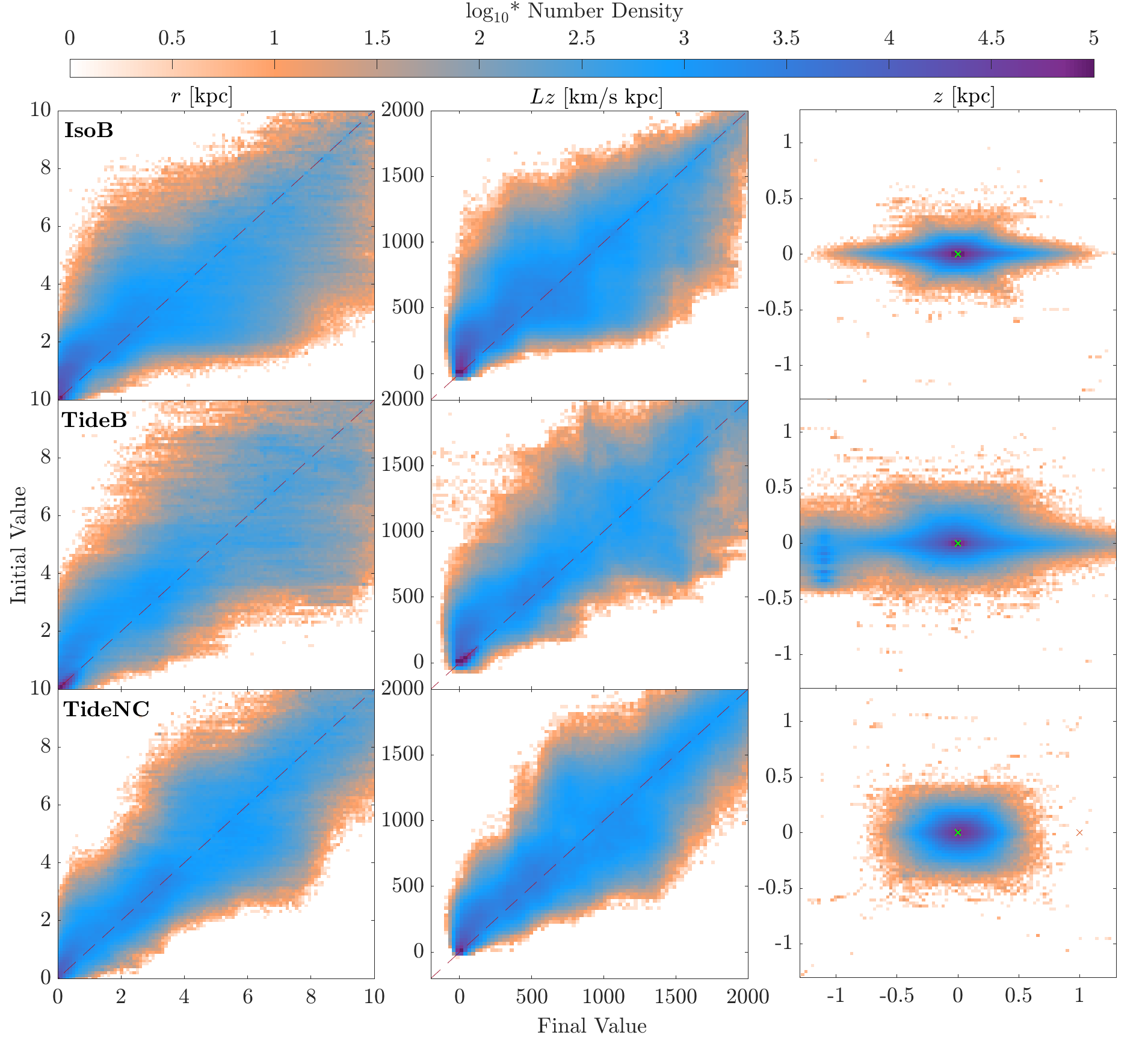}
  \caption{A representation of total change over stellar lifetimes (from star formation to 1\,Gyr) in $r$, $L_z$ and $z$ for stars formed in the simulations of each: IsoB, TideB and TideNC. Each value is divided into equal bins based on formation position and then separated into similar bins for the corresponding values in the 1\,Gyr snapshot. The colour is weighted logarithmically by stellar number density and a diagonal dashed-line represents the line of no change (initial=final). For the vertical component, where the expected response is dispersion about the disc-plane, a cross-marker indicates $z_i=z_f=0$\,kpc. } 
\label{f:initial-final}
\end{figure*}

To commence, a general assessment of how dynamical features change over the course of the lifetimes for stars formed in each of the three discs, the initial and final values for standard migration signatures: radial position $r$; scale-height $z$; and angular momentum $L_z$ over the 1\,Gyr simulation period are presented in Figure \ref{f:initial-final}. Stellar particles are binned based on the initial values, then the corresponding final values for each of the particles in each bin are subsequently counted into the same sized bins. This is then a representation of the maximum likelihood for how each attribute may change over the total allowed evolutionary period (1\,Gyr). A straight diagonal-line from bottom left to top right is drawn to indicate no overall change (as along this line the initial\,=\,final value). Spreading above this line would indicate a decrease in value over the simulation, while spreading below this line is evidence of the given value increasing with simulation time. This allows us to determine that change is indeed taking place in each of the key parameters and that the shape of these changes does initially appear to be influenced by differences in bar formation mechanisms and possibly even by bar-type or morphology. 

This is reflected in the similarities and differences exhibited by the three discs in each parameter. For example, the TideNC disc is most noticeably different from the two more strongly barred-discs (IsoB and TideB), with $r$ and $L_z$ values forming a tighter envelope about the line of no-change and, therefore, signifying a more consistent trend for these attributes to remain broadly unchanged over the disc evolution period. The distribution in $z$ is also almost circular around the point of $z_i=z_f=0$\,kpc. This is compared with the stellar components of IsoB and TideB that each exhibit significant vertical spreading for stars originally located within the disc-plane (at $z_i=0$\,kpc). The shape of the more significant deviations in $r$ and $L_z$ values, however, is not so consistent for the discs of IsoB and TideB. Both radial and angular momentum values in the IsoB disc appear furthest from the line of initial\,=\,final in the mid-regions, while TideB is increasingly spread towards the more extreme values (at IsoB: $r\sim2-6$\,kpc, $L_z\sim500-1500$\,km/s/kpc; TideB: $r\gtrsim4$\,kpc, $L_z\gtrsim1500$\,km/s/kpc). Interestingly, TideNC also shows the largest deviation from the line of initial\,=\,final in the mid-ranges of $r$ and $L_z$, similar to the other isolated disc (IsoB), despite the magnitude of this change being comparatively smaller (at $r\sim4-8$\,kpc, $L_z\sim1000-1500$). 

\begin{figure}
  \centering
	\includegraphics[width=0.98\columnwidth]{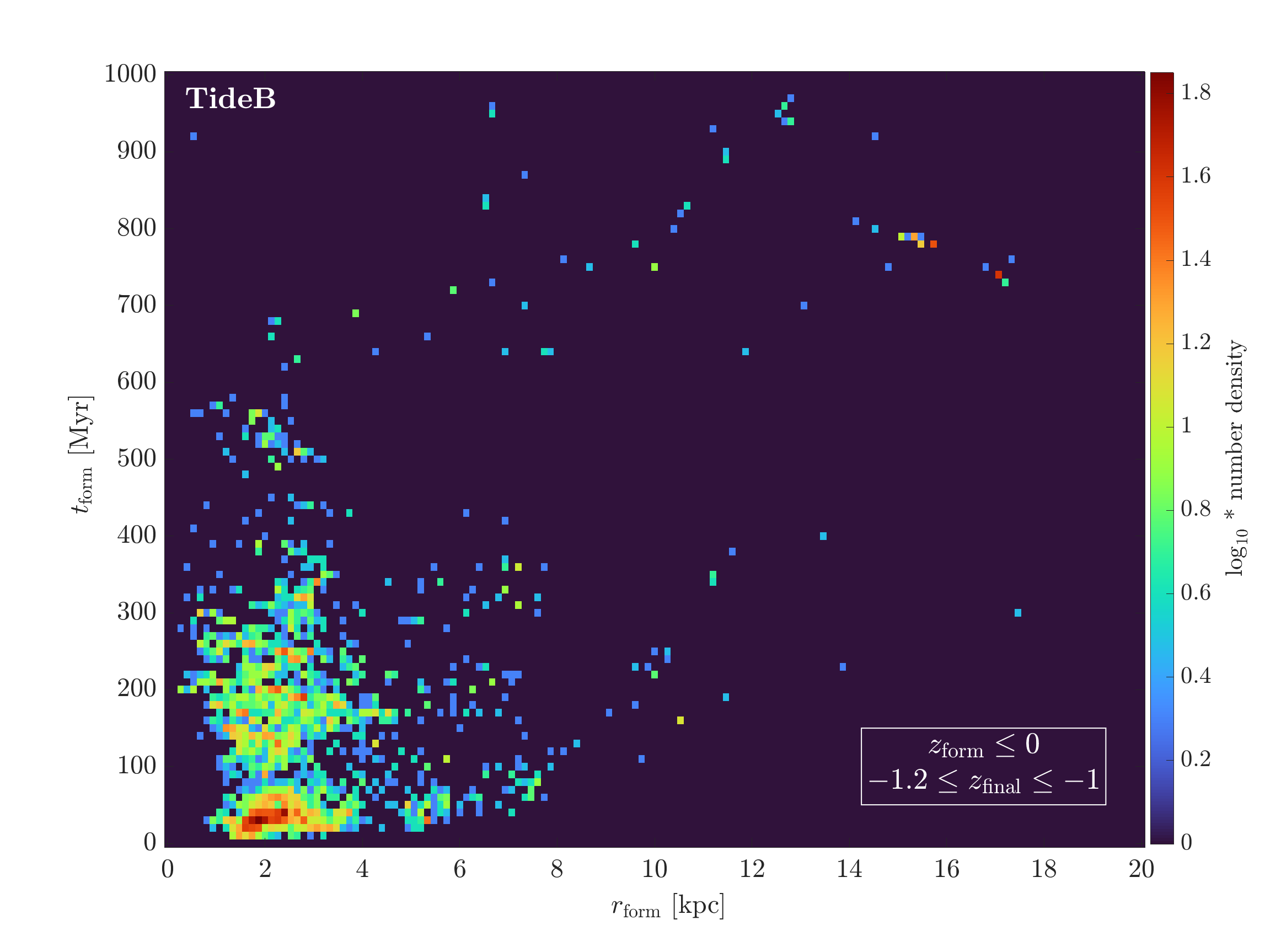}
  \caption{The number density of stellar particles in the TideB disc with initial values of $z<0$ and final values of $-1.2\le z\le -1$ with formation time and radial position information to account for the composition of the strange blob feature in the TideB $z$ panel of Figure \ref{f:initial-final}. } 
\label{f:form}
\end{figure}

The more significant flaring in the only tidally affected disc (TideB) can then be likely attributed to the interaction tidal forces more significantly impacting the stars on the outer-edges of the disc than the internal disc morphologies. Additionally, there is a significant fraction of stars with a large variation in initial vertical values ($-0.5\lesssim z\lesssim 0.5$\,kpc) which exhibit consistently large final positions below the plane ($z\gtrsim -1$\,kpc) forming a blob in the left of the TideB $z$ panel. These stars are predominantly formed before the period of closest-approach ($t_{\rm form}\lesssim 90$\,Myr) and with an additional significant component in the following 200\,Myr period ($100 \lesssim t_{\rm form}\lesssim 300$\,Myr). Interestingly, rather than being stars formed in the outer-radii of the disc most directly affected by the companion, these are mostly formed within the radii which would become the bar-edges ($1\lesssim r_{\rm form} \lesssim 4$\,kpc, centred on $\sim2$\,kpc, see Figure \ref{f:form}). Considering the formation times, however, it is still most likely that this must be a direct result of the interaction and could be worth further, targeted investigation. A discussion on the different effects of formation time and formation radius for radial migration of particles in the simulation is included in Section \ref{ss:form}. In a more general sense, the large elongation and deviation from spherical symmetry in the final values of $z$ for both TideB and the similarly barred IsoB, at the very least, indicates that the stars formed in these discs must experience significant vertical dispersion which is does not occur in TideNC. We would expect this to be related to the formation of the X-like feature in the galaxy cross-section which forms with the bars in IsoB and TideB but is not evident in TideNC (see Figure \ref{f:starprj_xzy}).

\begin{figure*}
  \centering
	\includegraphics[width=0.8\textwidth]{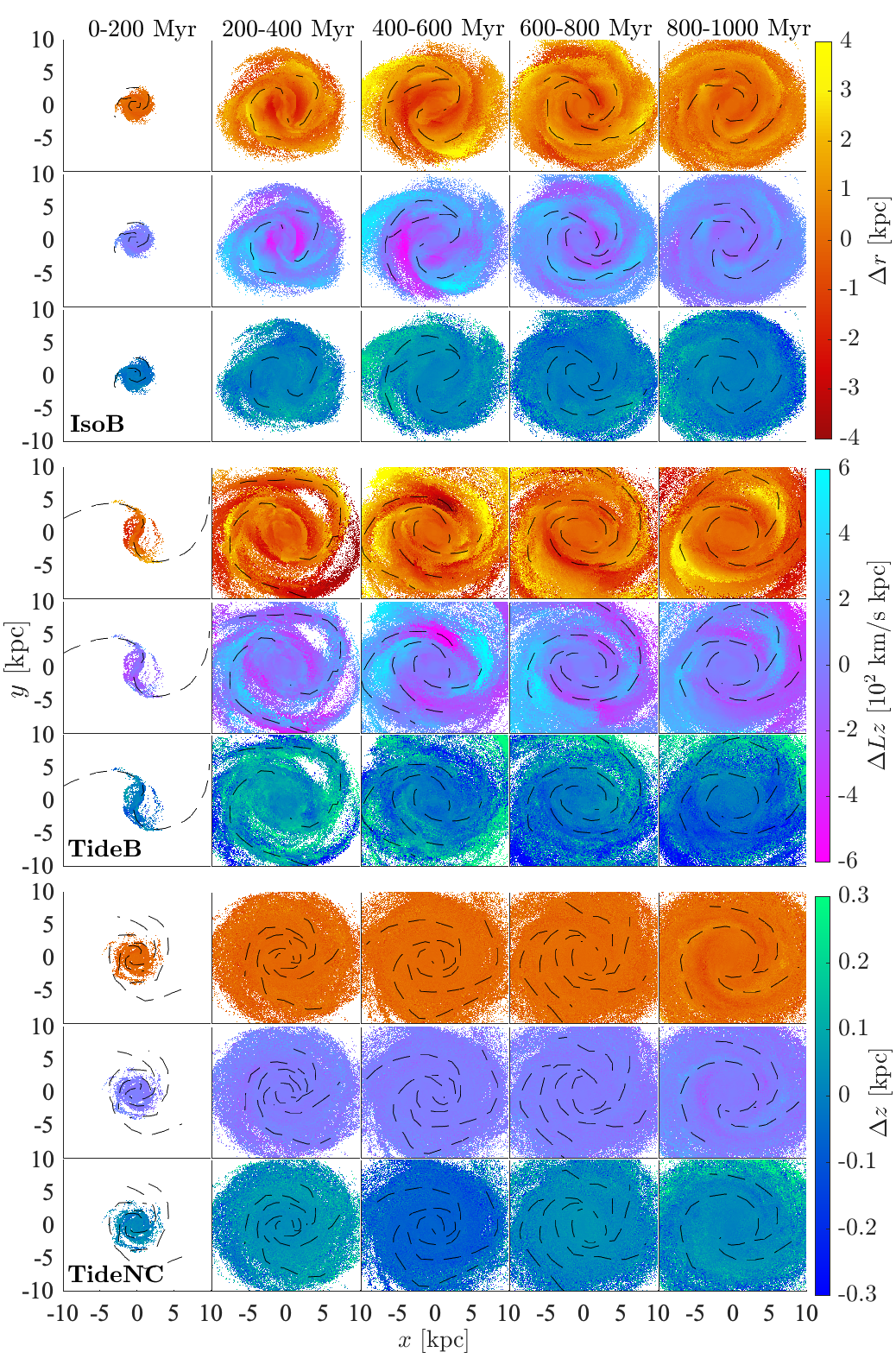}
  \caption{Projection of face-on changes in attributes $\Delta r$, $\Delta Lz$ and $\Delta z$ set into the $xy$-plane for newly formed stars in the simulation for each case: IsoB, TideB and TideNC. Median values in each grid square ($100\times100$\,pc) are used for colour weighting. Columns correspond to 200\,Myr periods of evolution while including only stars existing for the entire period. Black dashed lines trace arm-splines from the total stellar density in Figure \ref{f:starprj_xzy}.}
  \label{f:drdlzdzprj}
\end{figure*}

\begin{figure*}
	\includegraphics[width=\textwidth]{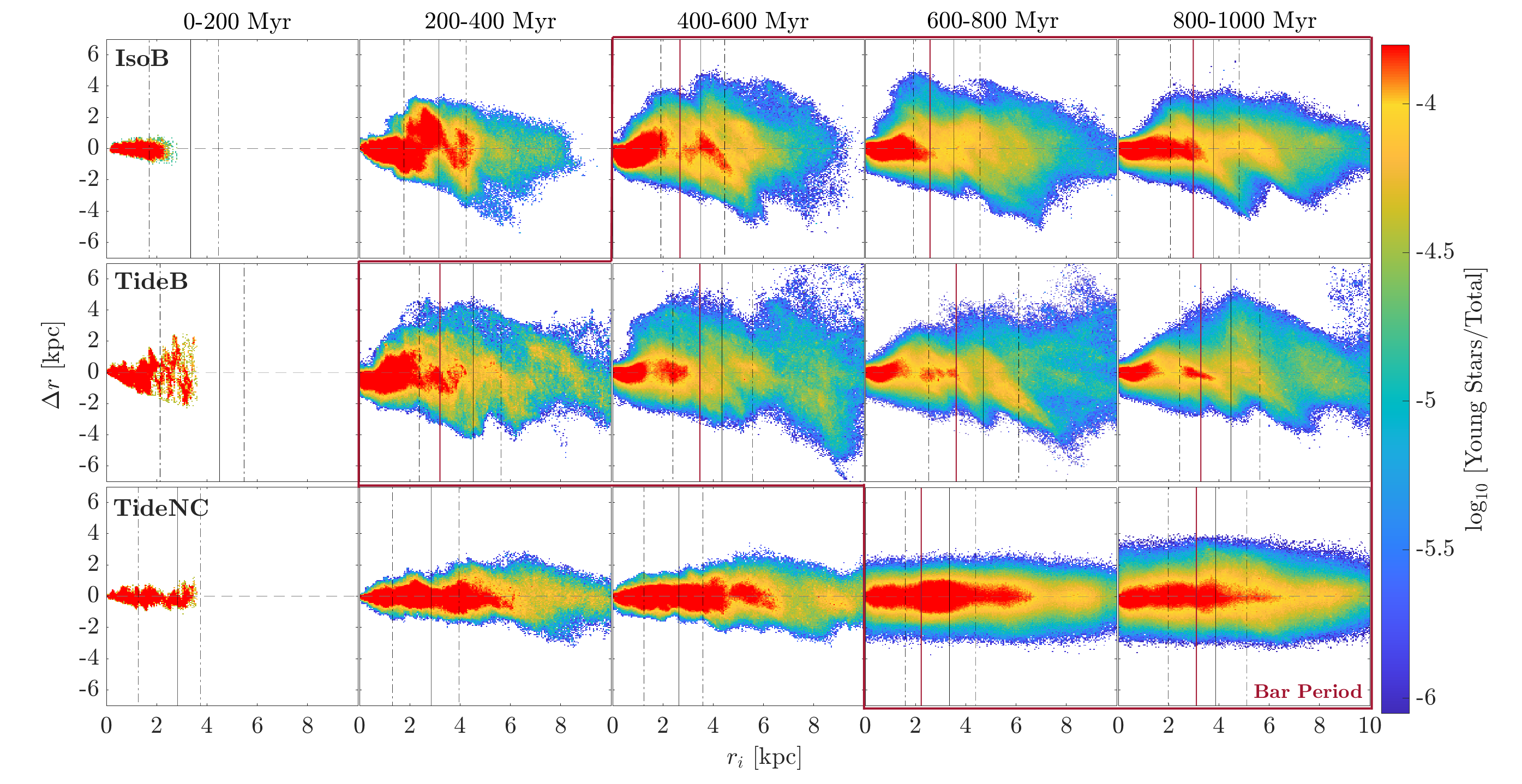}
  \caption{Dependence of attribute $\Delta r$ with radial position at a given period of evolution of stars formed in the simulation for each case: IsoB, TideB and TideNC. Columns correspond to 200\,Myr periods of evolution and include contribution from only stars present for the entire period to observe changes over the epoch completely. The colour is weighted logarithmically by stellar number density as a fraction of the total density while vertical lines trace bar-edges (red line), co-rotation (thin line), inner- \& outer-Lindblad resonances (dashed-lines).}
  \label{f:ridr}
\end{figure*}

These changes between initial and final values can be directly prescribed as a single change parameter for each attribute (i.e. $\Delta r$, $\Delta L_z$ and $\Delta z$). If assessed over the full sample of simulation formed stars within the 1\,Gyr iteration of disc evolution for each of the three discs, the results confirm the visual assessment from Figure \ref{f:initial-final}. The IsoB and TideB discs should evolve differently from TideNC with the total change for each $\Delta r$, $\Delta L_z$ and $\Delta z$ appearing as a more symmetric and Gaussian-like in distribution on average, centred on the most probable result of no change. While IsoB and TideB also show the most probable result is no change, both discs appear to indicate a preference for stars to undergo negative changes in $\Delta r$ and $\Delta L_z$ (i.e. inward radial migration and decreasing angular momentum) on average. Additionally, while there are larger magnitudes in changes for stars in TideB, there is also a higher probability of stars experiencing no change overall. This is not related to the different star forming efficiencies across the discs of IsoB and TideB \cite[e.g.][]{Iles2022}. So, stars in IsoB are comparatively more likely to experience change in position and angular momentum but these changes should not be as momentous as those observed in the tidally-driven TideB. 

To instead hone in on the correlation of migration patterns with face-on disc stellar structures, these changes in $\Delta r$, $\Delta L_z$ and $\Delta z$ are projected into the face-on $xy$-plane in Figure \ref{f:drdlzdzprj}. This is a map of the median change in each variable for the stellar particles within a $100 \times 100$\,pc grid square at the final time in each window of 200\,Myr evolution. Additionally, these values are only calculated for stellar particles which have already formed by the initial time of each window; for example, the 400-600\,Myr window contains contributions for only stars formed between 0 and 400\,Myr. This is to ensure that any changes which occur are measured over a consistent $\Delta t = 200$\,Myr. Obvious disc structures are reproduced in these projections, particularly in $\Delta r$ and $\Delta L_z$. The black dashed lines overlayed on this figure trace arm-spines in the stellar density from Figure \ref{f:starprj_xzy}. These are simple approximations traced by eye in a high-contrast logarithmic density image. However, from the overlay of these features it is possible to visually assess radii where the arms dominate often exhibit clearly defined structures with the largest changes in $\Delta r$ and $\Delta L_z$. However, this is also not always the case. Additionally, both positive and negative directions of change may be associated with the leading and trialing edges of the arms (i.e. $\Delta r < 0$ on the downstream side of the arms, as expected from previous studies). Arcs bracketing the bar, presumably from the arms decoupling, are particularly evident in both $\Delta r$ and $\Delta L_z$ for IsoB, and TideB in particular. TideNC, by contrast, has significantly weaker patterns in general, a likely consequence of the muted morphology compared to IsoB and TideB. It also appears that regions with large changes in $r$ similarly seem to exhibit large changes in $L_z$ with a corresponding consistency in the direction of these changes (i.e. large positive $\Delta r \Longleftrightarrow$ large positive $L_z$; large negative $\Delta r \Longleftrightarrow$ large negative $L_z$). There does not appear to be a similarly strong correlation between either $\Delta r$ or $\Delta L_z$ with features in $\Delta z$. Regions with the largest changes in $\Delta z$ appear to occur on the inner-edges of arms in the mid-outer parts of the disc, or just in the outer disc in general.

Based on this preliminary visual assessment, radial migration and angular momentum appear especially affected by the formation of disc structures, broadly tracing these non-axisymmetric features. However, changes in vertical position do not seem so strongly affected by identifiable disc-plane morphology on average. The TideB and IsoB discs, with similar bars, are more broadly similar, although there are noticeable differences, particularly in the magnitude of changes in the mid/outer disc regions, while TideNC appears significantly less prone to change overall.

\begin{figure*}
	\includegraphics[width=\textwidth]{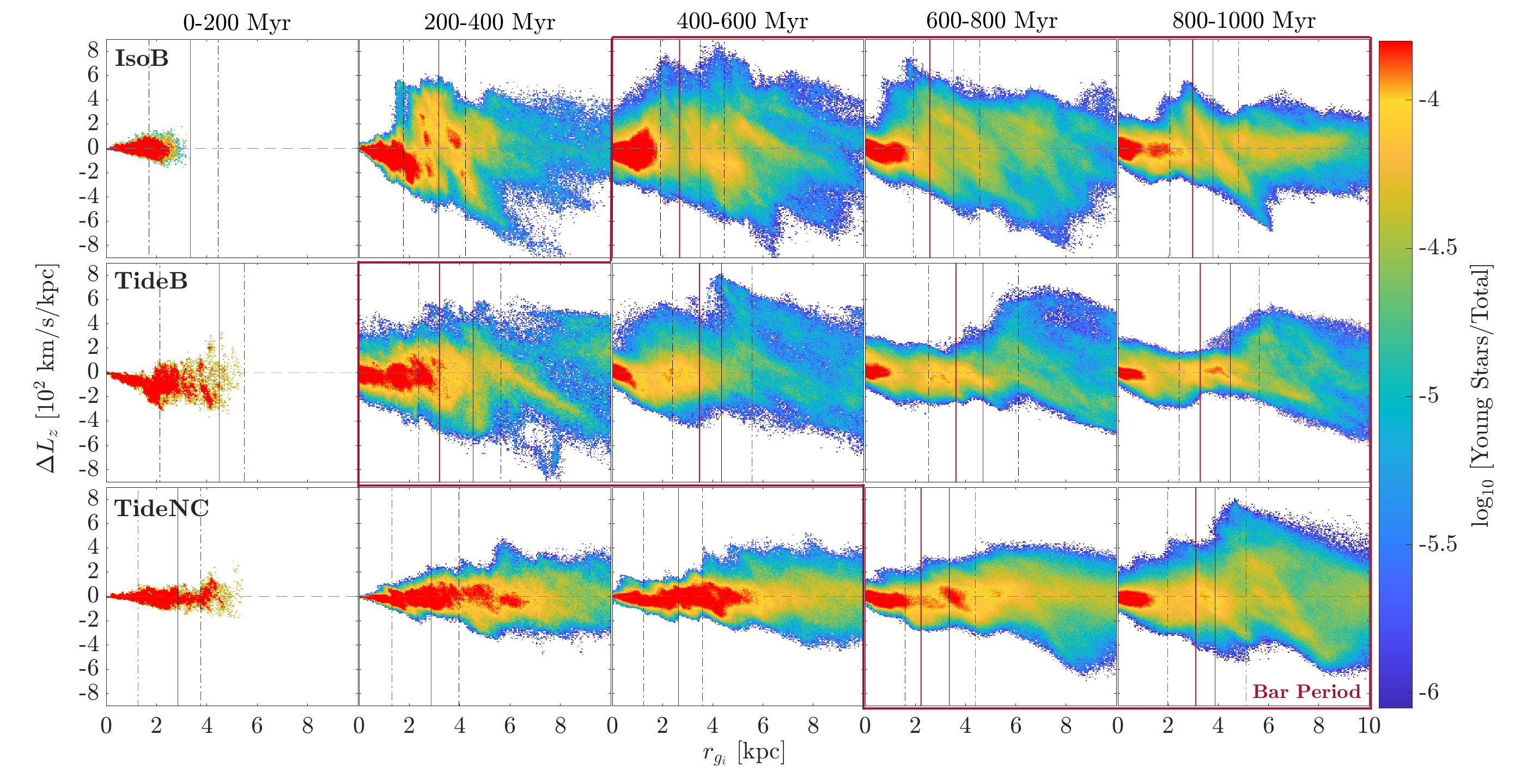}
  \caption{Dependence of attribute $\Delta L_z$ with radial position (defined by a guiding radius $r_g=L/V_0$) at a given period of evolution of stars formed in the simulation for each case: IsoB, TideB and TideNC. Columns correspond to 200\,Myr periods of evolution and include contribution from only stars present for the entire period. The colour is weighted logarithmically by stellar number density as a fraction of the total density while vertical lines trace bar-edges (red line), co-rotation (thin line), inner- \& outer-Lindblad resonances (dashed-lines).}
  \label{f:ridlz}
\end{figure*}

\subsection{Azimuthally-averaged Profiles}

In order to study the bulk migration properties of each disc, we calculate azimuthally-averaged profiles for the different orbital and kinematic properties of stars in each disc. We use initial radius ($r_i$) to prescribe the location where a star particle is located at the start of each period ($\Delta t = 200$\,Myr) wherein a change in parameter is measured. In this way, it is possible to constrain the migratory behaviour of stars from different sections of the disc. For all figures in this section, the colour weighting has been set to a fractional probability density where [\emph{number density of stellar particles to experience change}] / [\emph{total number of particles included in that period}], so that the different numbers of formed stars in each period of each disc will not affect the comparison. 

\subsubsection{Standard Migration Signatures}

Figure \ref{f:ridr} is a representation of the change in radial position ($\Delta r$) for simulation-formed stars in each of the three discs. Broadly, the shape of this distribution appears to vary considerably throughout the evolution of these discs. The responses of IsoB and TideB with similar bar features, exhibit more correspondingly similar traits, especially in the lead-up to and post-bar formation. Comparatively, TideNC appears to be consistent with the other isolated disc (IsoB) before the bars in these discs form, but is visibly different in latter periods post-bar formation. 

Significant diagonal spreading, like smudging sections of the envelope, appears consistently in the responses of all three discs but the magnitude and radial location of these features varies. IsoB exhibits one such feature centred on $\sim4$\,kpc with perhaps a second, smaller feature at about 6\,kpc which is evident in all epochs (seen in four right-most panels). By visual inspection, the centre of these smudges is often closely aligned with significant resonance positions, such as co-rotation and the inner- or outer-Lindblad resonances. These are derived from pattern speeds determined by direct particle tracing as in previous studies of these discs \citep{Iles2022} -- see vertical lines delineating resonances (CR: black, solid; ILR, OLR: black, dashed) and bar-edges (red, solid). Although the alignment of these smudges does not always appear to be exactly co-located with such resonances, the number and radial locations of these features appear relatively constant over the disc evolution in IsoB. Comparatively, TideB features similar diagonal spreading throughout the disc and yet, the number and radial location of these features is less consistent over time. The alignment with the resonance positions is also not as clear, although these are not necessarily un-aligned (see, for example, the $400-600$\,Myr window, where yellow smudge features intersect the resonance positions clearly). On the other hand, TideNC does not show such significant diagonal smudging. There is some evidence of warps in the envelope in early periods (see $200-400$\,Myr) but by the time the bar forms the features in this plot have become particularly smooth (see two right-most panels). Although, there does appear to be some small variations in the shape of the density distribution within this consistent envelope--similarly almost-aligned with co-rotation and the other resonance positions. It is evident that even the disc resonances do not seem to elicit similarly strong migratory responses from the stellar populations of TideNC, particularly post-bar formation when compared to the other two discs with noticeably stronger bar structure. 

\begin{figure*}
	\includegraphics[width=\textwidth]{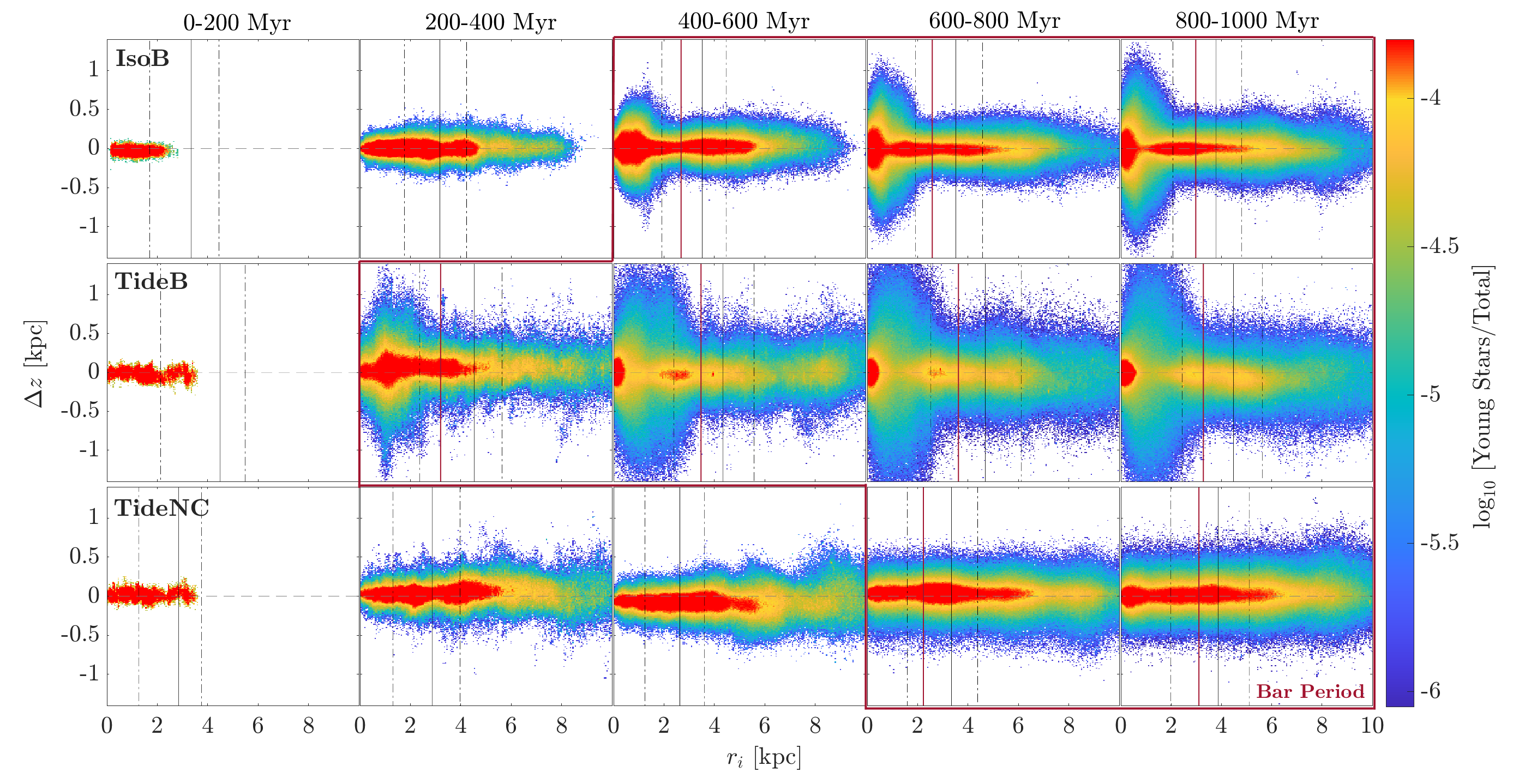}
  \caption{Dependence of attribute $\Delta z$ with radial position at a given period of evolution of stars formed in the simulation for each case: IsoB, TideB and TideNC. Columns correspond to 200\,Myr periods of evolution and include contribution from only stars present for the entire period. The colour is weighted logarithmically by stellar number density as a fraction of the total density while vertical lines trace bar-edges (red line), co-rotation (thin line), inner- \& outer-Lindblad resonances (dashed-lines).}
  \label{f:ridz}
\end{figure*}

Additionally, the most likely value for all three discs appears to be approximately zero on average (red coloured: $\ge -4$\,dex) with IsoB and TideB slightly indicating a preference for small inward migration within the central region (see fraction of red component below horizontal dashed line), particularly at and just preceding bar formation (i.e. $200-400$\,Myr in IsoB \& TideB; and $400-600$\,Myr in IsoB). In the region of $r_i \lesssim 1.5$\,kpc most stars in IsoB and TideB appear to favour a predominantly inward trend ($\Delta r < 0$), with this inward trend increasingly in magnitude as initial radii increases. However, for slightly larger radii ($r_i \gtrsim 1$\,kpc) this trend appears to equalise, or even reverse, as a much higher numbers of stars seem to also favour similar magnitudes of outward migration as $r_i$ increases towards the bar-edges. This is perhaps evidence of centrally formed star particles being preferentially trapped in the nuclear stellar disc (NSD) region whereas migration within the wider disc is much more regular. There is also a notable feature in TideB where this most likely (red) value shows a clear separation for stars originally located at radii just inside the barred region, which is not replicated in either of the isolated IsoB or TideNC. This indicates that there is either some active force preferentially driving stars to leave from these radii; or, that there are significantly fewer stars in this region compared to the surrounding radii and yet, a similar fraction of the total number of stars overall will still move away. This separation also appears to be close to the radii at which the preferentially inward trend of centrally located stellar particles may equalise and may also be related to the presence of a NSD. 

Similar features are present in the change in angular momentum ($\Delta L_z$), with the smudge-like features appearing even more distinct. Many diagonal smudge-like features identified in $\Delta r$ for both IsoB and TideB are evident at the same $r_i$ for $\Delta L_z$, however, as angular momentum is related to both position and velocity components, we also produce an alternate, correspondingly momentum dependent, radial measure to analyse these features. By using the initial guiding radius $r_{g_i} = L/V_0$ with $V_0$ prescribed by the rotation curve at the initial time period for each epoch of the simulation, rather than simply the initial radius, features in $\Delta L_z$ are more distinct (see Figure \ref{f:ridlz}). The other parameters identified in this section (e.g. $\Delta r$, $\Delta z$) were also tested against this guiding radius and produce no obvious differences thus, the assumption of such a parameter with a dependence on angular momentum seems most appropriate to assess the change in angular momentum herein.

Figure \ref{f:ridlz} also exhibits large smudge-like features, mostly centred on or around co-rotation and the outer-Lindblad resonance for IsoB and TideB with less significant features in TideNC (see, for best example, the black solid-line (CR) in $800-1000$\,Myr window of IsoB or the black dashed-line (OLR) in $200-400$\,Myr of TideB). However, compared to the almost featureless appearance of TideNC in terms of radial change $\Delta r$, there are many more obvious features in the barred periods of this disc in $\Delta L_z$, particularly around co-rotation and in the outer-disc. Additionally, an evolution of these features in angular momentum appears to occur more significantly over the total simulation period. For IsoB, only the inner ($\sim$4\,kpc, co-rotation) feature remains particularly strong in latter periods. Comparatively, the mid-disc feature ($\sim6$\,kpc, outer-Lindblad resonance) in TideB remains most prevalent by 1\,Gyr with the magnitude of change in momentum within the inner regions decreasing over time, particularly around the bar-extent. Where change is occurring in both radial position and angular momentum, this indicates that churning should be the most likely driving mechanisms for the stellar migration; that is, torques from non-axisymmetric features in the disc should be primarily affecting the stellar motion \citep[e.g.][]{Sellwood2002}. Considering that most of these consistent features are closely aligned with the disc resonance positions, this is perhaps significant. Most notably, while the inner-most regions of both IsoB and TideB ($r_i \lesssim 1.5$\,kpc) previously showed a similarly increasing tendency toward negative values with increasing radius, this is not reproduced as strongly in either disc for $\Delta L_z$ with both discs showing much more symmetric results in this region. The only exception is the $400-600$\,Myr period of TideB, although it remains unclear what may have cause this predominantly decreasing angular momentum for $r_{g_i} \lesssim 1.5$ in this period alone. The odd separation in the most likely (red) value also remains persistent but to a significantly lesser extent. Again, this does not appear to be replicated in either IsoB or TideNC nor is the location of this feature consistently co-located with any of the significant resonances or smudge-like features. 

\begin{figure*}
	\includegraphics[width=\textwidth]{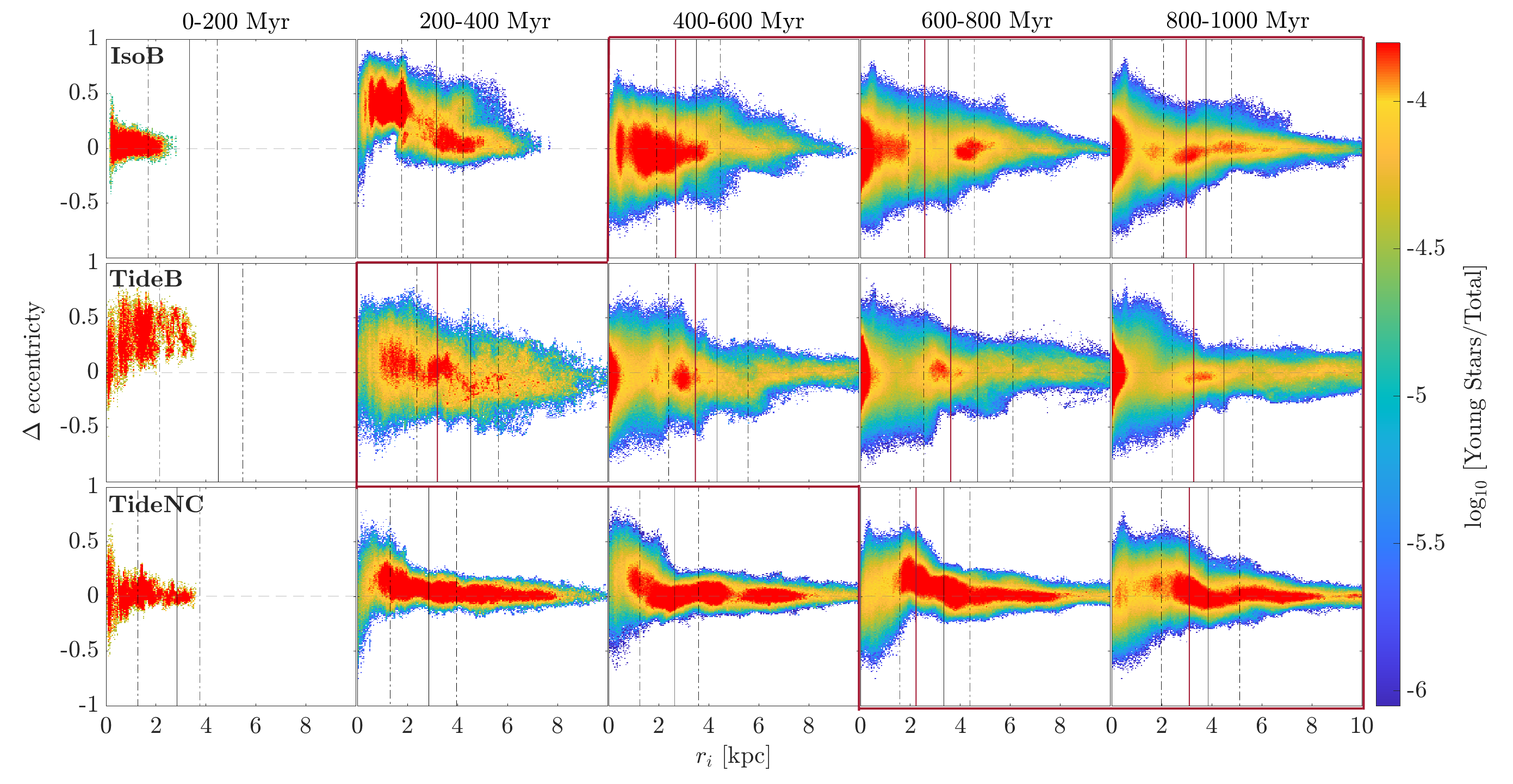}
  \caption{Dependence of attribute $\Delta {\rm{ecc}}$ with radial position at a given period of evolution of stars formed in the simulation for each case: IsoB, TideB and TideNC. Columns correspond to 200 Myr periods of evolution and include contribution from only stars present for the entire period to observe changes over the epoch completely. The colour is weighted logarithmically by stellar number density as a fraction of the total density while vertical lines trace bar-edges (red line), co-rotation (thin line), inner- \& outer-Lindblad resonances (dashed-lines).}
  \label{f:ridecc}
\end{figure*}

Considering the change in vertical position (Figure \ref{f:ridz}), some indication of what may be causing these stars to preferentially leave this radial location appears. The formation of the bar in IsoB and TideB can be seen to significantly drive vertical motion both above and below the disc plane. This is in accordance with literature on bar formation, wherein bars have been shown to warp in and out of the plane during the early stages of formation before settling into secular evolution \citep[e.g.][]{Friedli1993, lokas2014, Sellwood2020}. Large changes in $z$ position occur for precisely the barred periods of IsoB and TideB. A similar arrow-head shape is formed within the bar region with the arrow-point at ($r_i$,$\Delta r$)=(0,0) and the maximum spread just inside the bar extent (red solid-line). The maximum extent in $\Delta z$ of this flaring in TideB is almost twice that of IsoB (e.g. $\sim$1.2\,kpc compared to $\sim$0.6\,kpc in the 400-600\,Myr epoch). The strange separation in most likely value from the previous analysis of $\Delta r$ also occurs when the spread of $\Delta z$ is greatest which should suggest that these stars are moving significantly in the vertical ($z$) and radial ($r$) directions. This feature may then be attributable to the formation of the X-shape seen in Figure \ref{f:starprj_xzy}. Therefore, there must be some vector component force moving stars more significantly out of the standard orbits within the disc-plane and more significantly than stars in the surrounding regions (i.e. a large vertical change and more significant inward or outward radial motion are strongly correlated for this specific region within the tidal bar). It is considered possible that this may be related to vertical resonances rather than the disc-plane resonances \citep[e.g.][]{Pfenniger1991} or, at least, the combination of these resonances acting in conjunction with the the disc-plane resonances. 

Additionally, while the face-on stellar population for latter periods of TideNC also exhibits bar-like morphology in the central region, there is no indication of similar vertical spreading in $\Delta z$ at any epoch in Figure \ref{f:ridz}. This is consistent with the lack of vertical (cross- or X-like) features in the side-on projection of TideNC (see Figure \ref{f:starprj_xzy}). As no significant vertical motion is occurring in TideNC, no obvious vertical structure is formed. Some small wiggles can be seen in the envelope at various disc radii in earlier periods but within the barred epoch, the envelope of TideNC is generally flat ($\Delta z = \pm 0.5$\,kpc) and only a small bulge or wiggle is visible in the most likely (red) value of the density distribution within this envelope. This indicates that the change in vertical position experienced by all stellar particles in this disc must be generally consistent, regardless of initial radial position and despite all apparent non-axisymmetric features in the disc-plane. 

\subsubsection{Orbital Parameters}
In addition to changes in stellar positions and momenta, we also consider the effect of structure evolution on the eccentricity of the stellar orbits in each disc. These orbits are defined for each star particle formed in the simulations via approximating the minimum and maximum radial positions ($r_{\rm{min}}$ and $r_{\rm{max}}$) within a given time period as analogues to the semi-major and semi-minor axes of an elliptical orbit. By convention, the eccentricity under this approximation is prescribed to be ${\rm ecc} = (r_{\rm{max}}-r_{\rm{min}})/(r_{\rm{max}}+r_{\rm{min}})$ with a calculation period of 100\,Myr. This allows for the determination of a variable $\Delta {\rm ecc}$ calculated over 200\,Myr periods comparable to the previous analysis. However, it is noted that this is not the most accurate choice for such an approximation of orbital parameters. Based on the rotation speeds, this 100\,Myr may only be sufficient to capture a complete revolution of the disc in the central regions ($r\lesssim3$\,kpc) and so the minimum and maximum values found for mid/outer-disc particles may not be the true values for a complete circular orbit. Thus, it is important to note that the precision of these measurements decreases with increasing disc radii. However, this value still functions as a practical approximation for the changes in apparent eccentricity of the stellar orbits, relative to initial radial position, over short developmental time periods for structure formation as in Figure \ref{f:ridecc}. 

\begin{figure*}
	\includegraphics[width=0.75\textwidth]{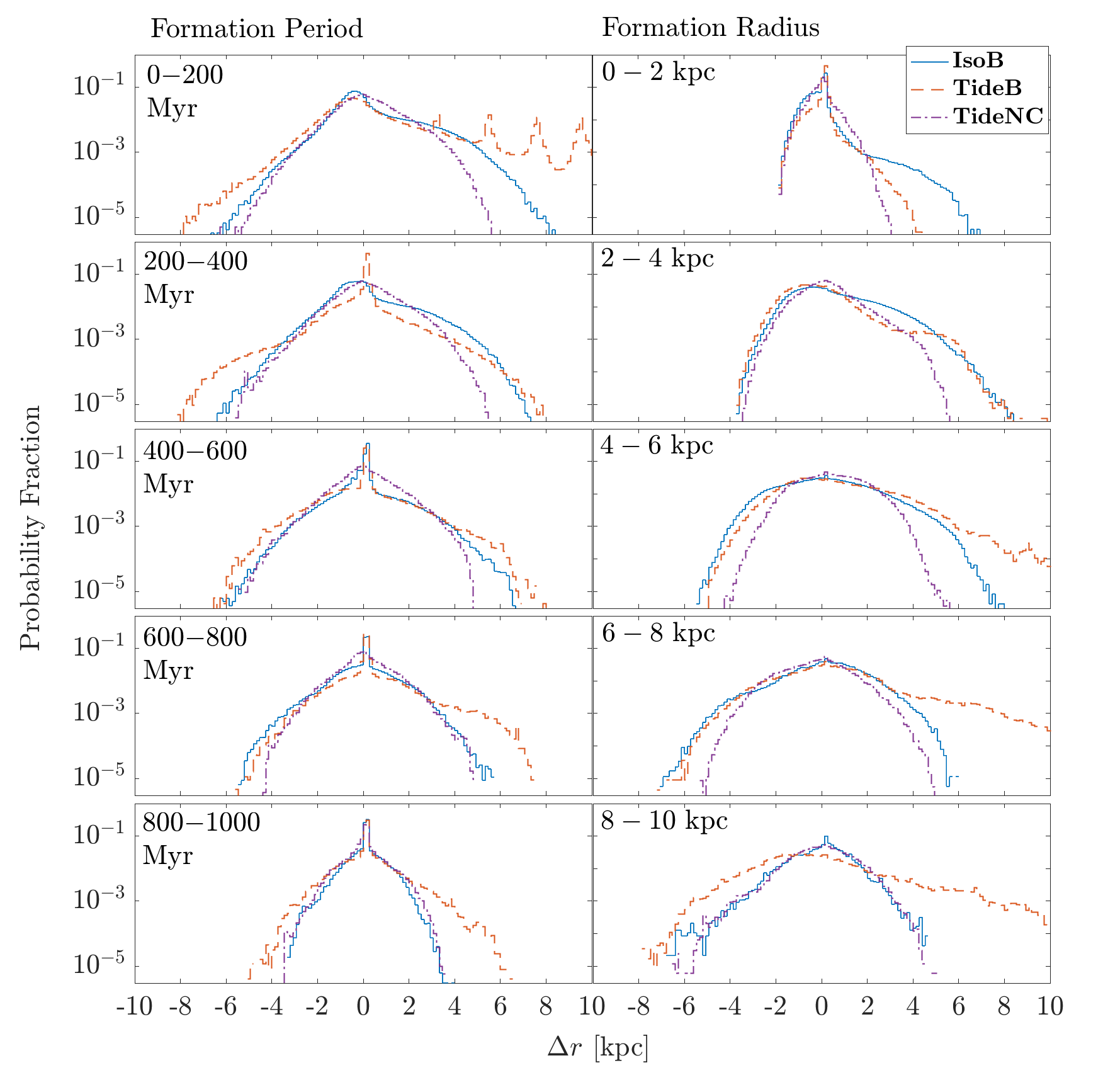}
  \caption{A histogram representing the total change over the simulation time (1\,Gyr) in $\Delta r$ for stars formed in the simulation for each: IsoB (solid blue line), TideB (dashed orange line) and TideNC (dot-dashed purple line). The stellar population is separated into five bins based on formation time in generations of 200\,Myr (left) and formation radius in radial annuli of width 2\,kpc (right).}
  \label{f:hist_drdlzdz_perage}
\end{figure*}

From Figure \ref{f:ridecc}, it appears that bar formation is generally preceded by large positive changes in the eccentricity of orbits for stars within the soon-to-be barred region. In fact, in most cases this feature is notably contained within the radial location of the inner-Lindblad resonance. The feature is most obvious in the panels preceding the barred periods of both IsoB ($200-400$\,Myr) and TideB ($0-200$\,Myr) and, although a similar shape is present in periods of TideNC (for windows $200-400$ to $600-800$\,Myr), the overall positive values of change to eccentricity in IsoB and TideB are much less significant overall (closer to zero on average) and much more prolonged ($\sim 3 \times$ longer) in TideNC. After the bar has formed, however, both IsoB and TideB with similar bars are also similarly represented in the eccentricity profile. In the central regions, eccentricity is most affected (largest possible values in $\pm \Delta {\rm ecc}$) and this decreases towards the outer-disc (seen by a narrowing of the profile). While there are some small perturbations to the profile, particularly around the bar-ends and resonance positions, this trend is generally consistent. Interestingly, this indicates that the interaction which may still be affecting the outer-disc of TideB is not significantly impacting the eccentricity of these outer-disc stellar orbits past the period of closest-approach (at $\sim100$\,Myr which is within the $0-200$\,Myr window, where the profile of $\Delta {\rm ecc}$ remains wide at larger radii).

\subsection{Formation Radius \& Formation Time}
\label{ss:form}
The previous analysis demonstrates that differing bars and bar formation mechanisms may indeed differently affect the stellar motion within galaxies. However, simulations have the advantage of being able to sample a complete picture of galactic evolution, which is not possible for real observations. As such, we now restrict analysis of the stellar discs to a single snapshot (after 1Gyr) to investigate possible signatures of these bar features in observable stellar populations.

\begin{figure*}
	\includegraphics[width=1.03\textwidth]{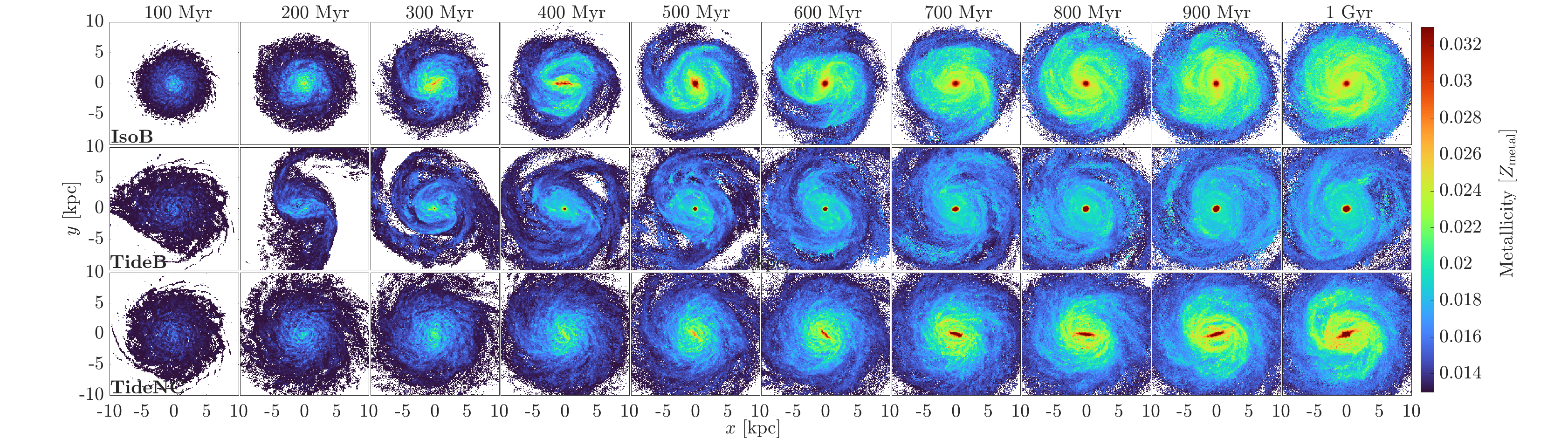}
  \caption{Projections for the face-on stellar metallicity distributions in terms of simulation metallicity parameter $Z_{\rm{metal}}$ set into the $xy-$plane for each case: IsoB, TideB and TideNC. Columns step in time by 100\,Myr to the total simulation time of 1\,Gyr.}
  \label{f:ZmetalPrj}
\end{figure*}

Figure \ref{f:hist_drdlzdz_perage} is a histogram to represent the total radial changes ($\Delta r$ from formation to current time) for stars in the 1\,Gyr snapshot, separated based on either the formation period (left column) or formation radii (right column). In the left panel of this figure, the stars formed in the simulation are separated into populations based on the stellar ages of particles. This displays how the probability of a star moving from its formation radius to the current observed radius, varies for the different populations of stars. Over time, the width of these features in the left column, appear to narrow for all three discs--that is, the youngest population of stars exhibit smaller changes in radius overall compared to older stars. This is not altogether unexpected as these younger stars have had less total time to move from their formation positions. Additionally, the TideB stars experience the largest changes in radial position of all three discs, particularly those stars formed in periods around the closest-approach of the companion, which occurs at $\sim100$\,Myr (i.e. the 0-200\,Myr population) and drives the bar formation in this disc. However, these comparatively larger changes in radius are evident in all populations of TideB, indicating the effect of the interaction remains noticeable even two rotation periods after closest-approach with even younger stars migrating considerably more than in the other (isolated) discs. The IsoB stars formed in earlier periods (0-600\,Myr) also appear to experience comparatively greater changes in position overall than stars formed at a similar time in TideNC, although less than in TideB. This trend, however, is not replicated for the youngest stars, as can be seen from the 800-1000 Myr panel. These stars appear to experience almost exactly identical changes in radial position in both IsoB and TideNC, the discs with isolated evolutionary scenarios post-bar formation. 

In contrast, the histograms presented in the right-hand side of this figure are separated based on the radial positions where these stars were formed, irrespective of when the star formation took place. As could be expected, there is an exceedingly high number of stars formed in the outer regions ($r_{\rm{form}} \ge 6$\,kpc) of the TideB disc which experience the maximum values of outward migration. This is very likely the consequence of the companion stripping gas and stars away from the disc. This trend toward higher positive values for $\Delta r$ in TideB is also evident to a lesser extent for stars formed between 4-6\,kpc, while stars formed between 2-4\,kpc clearly experience similar changes overall to the IsoB disc with similar bar morphology. For stars formed within the centre-most region, however, TideB is significantly less likely to produce the same extent of radial change as IsoB. Stars formed in all three cases are more likely to experience small inward or zero changes overall but the probability distributions for IsoB and TideB are significantly more asymmetric than TideNC. Stars formed in the central and mid-regions of these two discs ($r_{\rm{form}} \le 6$\,kpc) are skewed as these stars cannot produce a $\Delta r < -r_{\rm {form}}$, leading to a long probability tail extending to larger positive values and indicating outward migration. This continues to be true for the outer-radii of TideB as mentioned previously; however, IsoB is instead slightly skewed to the opposite direction for formation radii of $r_{\rm{form}} \ge 6$\,kpc, with an asymmetrically more extended distribution to negative $\Delta r$ values. Interestingly, the last panels of both columns are similar in that the IsoB and TideNC probability distributions appear to align only in this window. For the separation based on formation radius, this means that the behaviour of stars formed in the outer-edges of the disc ($r_{\rm{form}} \ge 8$\,kpc) is similar for both isolated discs (IsoB \& TideNC) without the tidal interference of the companion, as could well be expected.

In summary, in the tidally driven disc, the extent of migration is heavily dependent on stellar age (or the periods when stars were formed, relative to closest-approach), as well as where (radially) in the disc stars were formed. Comparatively, both isolated discs develop similarly. The extent of the migration in each isolated disc appears generally irrespective of stellar age. However, the radial position at star formation does appear important for determining the migratory tendencies of stars in isolated discs, similar to the tidal condition. Despite this apparent similarity, these trends in the radial dependence at star formation for the isolated discs do not appear qualitatively related to those in the radial dependence at star formation for the tidal disc. Hence, this could become a distinguishing trait of the two mechanisms.

\subsection{Metallicity Distribution}

Observations of stellar metallicity are often used for tracing stellar population mixing and resolving many significant features pertaining to the evolutionary histories of galaxies \citep[e.g.][]{Molla1997, Haywood2013, Magrini2009, Hayden20, Lacerna2020}. Hence, we attempt to perceive differences and features within the metallicity distribution at the final 1\,Gyr period of the three simulated discs (IsoB, TideB \& TideNC). However, the metallicity in these simulations should be viewed while taking into consideration that all gas in the simulation was initially given a Solar value of $Z_\odot = 0.013$ for convenience. While, physically, a metallicity gradient appears ubiquitous in observations of real galaxies, for the purpose of simulation, it is unclear exactly which precursor metallicity profiles have developed into these observed gradients. By setting metallicity to initially be constant, any changes in the distribution are easy to measure by way of the variance from this pre-set value (in this case, the Solar metallicity). However, due to such an idealized initialization, these discs lack a fully self-consistent treatment of stellar metallicity comparable to observational values. See \citet{Stinson2006} for a more thorough description of the metallicity routines used in \textsc{Gasoline2}, along with \citet{Wadsley2004, Shen2010, Wadsley2017} for other code specific and sub-grid prescriptions which may also be relevant. Hence, while simulations evolved in this way should be assessed with caution in comparison to true observed values, it is deemed sufficient for a preliminary, relative assessment of disc metallicity features, perhaps via normalised metallicity measurements. Differences in the treatment of on-the-fly metallicity evolution and a study of more complex initial metallicity profiles for simulations such as these, would certainly be an area to explore in future works. 

Figure \ref{f:ZmetalPrj} shows the evolution of stellar metallicity for the three discs over the 1\,Gyr period. The two similarly barred discs (IsoB \& TideB) form an obvious circum-nuclear region with the highest metallicity stars almost as soon as the bar forms (see Figure \ref{f:starprj_xzy}). In comparison, the bar feature of TideNC is highlighted by a long, thin strip of high metallicity stars in the central regions of this disc, even long before the bar is obviously formed in the stellar structure. A similar feature is also briefly evident in IsoB during the pre-bar period (300 and 400\,Myr panels), which may be a sign of a bar forming in isolation. The central circular shape is, however, notably absent in TideNC. Additionally, the interaction in TideB appears to drive a significantly higher ratio between the central (nuclear) metallicity and the mid- to outer-disc metallicity, when compared to the discs of the two isolated bars which appear relatively similar in this aspect (IsoB, TideNC: green/bright disc structures $\mathcal{O}$[0.02-0.03]; TideB: bluer/fainter disc structures $\lesssim 0.02$). However, despite the difference in magnitude, all three of these discs are similar in that higher metallicity regions trace the major disc structures, such as the bar and arms. This could make identifying this feature, which is inherently comparative, difficult in a single galaxy observation. We also note that this is not simply related to differences in the star formation rates (SFR) in these disc regions. While there is a difference in the number of stars forming in the inner-regions of these discs compared to the mid-/outer-regions and this does seem to correlate in general, it is likely not a direct consequence. In a comparison between the two isolated discs, where the ratio of stars formed in the inner ($\le2$\,kpc) to outer ($>2$\,kpc) disc is significantly different (IsoB: $\gtrsim 1$; TideNC: $\lesssim 0.5$), the median metallicity ratio from inner-outer disc is almost exactly equal (IsoB: $\sim 1.90$; TideNC: $\sim 1.89$). The tidal disc is also distinguished by distinctly higher values for both these attributes (ratio: SFR $\sim 1.40$ ; $Z_{\rm metal} \sim 2.44$). 

There have been many studies, both numerical and observational, attempting to trace and analyse the differences in metallicity in both radial and azimuthal (tracing arms) features of disc galaxies. Naturally, with higher values of star formation occurring in the arms than the inter-arm regions of the disc \citep[see][for IsoB \& TideB]{Iles2022}, it is not surprising that it is possible to observe higher metallicities tracing these features in the projections of the three discs in Figure \ref{f:ZmetalPrj}. Some azimuthal variation in the metallicity is also to be expected, due to streaming motions of gas along spiral arms, which has been detected in a number of similar observed spiral-galaxies, such as NGC\,0628, NGC\,1087, NGC\,1672, NGC\,2835 and NGC\,6754 of which all but NGC\,0628 are classified as barred \citep[e.g.][]{Sanchez-Menguiano2016, Kreckel2019}. Comparatively, while many previous studies agree that the presence of a bar should drive gas inflows towards the galactic centre \citep[e.g.][]{Athanassoula1992, Kormendy2004, Jogee2005, Wang2012, Cole2014, Baba2020} thus, transporting metal-poor gas from the outer-disc inwards, observations do not consistently reflect systematically flatter metallicity gradients for barred-galaxies on the whole \citep[e.g.][]{Vila-Costas1992, Martin1994, Sanchez-Blazquez2014, Kaplan2016, Kreckel2019, Li2022}. However, it is interesting that a number of these studies indicate that specific attributes, such as the strength of the bar or whether a galaxy is interacting, may more consistently appear to correlate with the flatness of the metallicity gradient observed \citep[e.g.][]{Kreckel2019, Li2022}. As bar strength and interaction history are precisely the main differences between the three discs showcased herein, these observations may indicate it could indeed be possible to observe the differences in metallicity across the disc highlighted previously (from Figure \ref{f:ZmetalPrj}) by determining some metric to describe the flatness of the disc metallicity gradient.

\begin{figure}
	\includegraphics[width=\columnwidth]{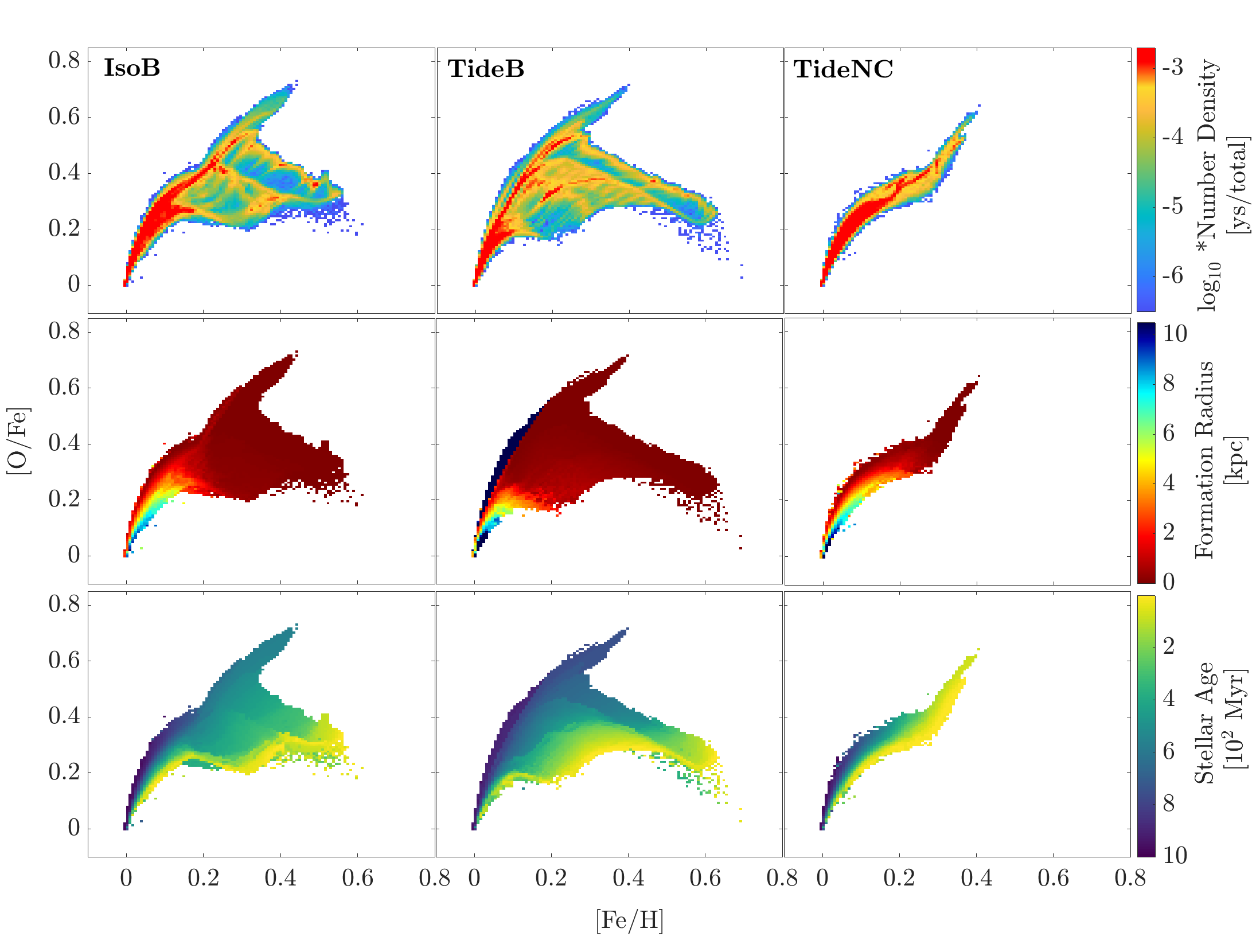}
  \caption{The fractional elements ([O/Fe] and [Fe/H]) at 1\,Gyr for stars formed in each of the three discs: IsoB, TideB, TideNC. The upper panel is weighted logarithmically by fractional stellar probability density; the middle panel by formation radius, with blue = outer-disc to red = inner-disc; and, the lower panel by stellar age, with youngest = lightest (yellow) and oldest = darkest (navy blue). }
  \label{f:OxFe}
\end{figure}

With the inherent benefit of observations to trace complete stellar histories, it is also possible to use these simulations to trace specific metallicity features of the \textquoteleft current\textquoteright \ (1\,Gyr) distribution, while simultaneously, identifying the different contributions of stars formed at different radii or with different ages. For example, relative abundance features are presented in Figure \ref{f:OxFe} for the fraction of elements [Fe/H] and [O/Fe] while independently weighted by the fractional stellar number density (similar to previous figures; upper), the radial position of stars at the time of formation (middle) and the age of each stellar particle (lower). However, we note that trends in this simulation are affected by the initially assumed value of [O/Fe]$=0$ which means these results only trace chemical evolution from relatively recent star formation and the [O/Fe] values are generally analogous to the SFR in regions of the simulated discs rather than true abundances. Nevertheless, a gradient from inner- to outer-radii corresponding to metallicities with high to low values for the [O/Fe] fraction is evident, with this feature significantly shallower for the interacting TideB than either of the isolated discs (IsoB and TideNC). Additionally, TideB has a component of stars formed in the outer-disc ($r_{\rm{form}} \ge 9$\,kpc) which occupies the left-facing edge of the metallicity distribution. This is not seen in either of the isolated discs. This component is clearly comprised of the most early formed stars and is expected to be directly related to the disruption in the outer-disc of TideB as the companion passed at closest-approach ($\sim$100\,Myr). This edge is also more linear in the TideB result, tracing a smoother curve to higher fractions of [O/Fe] for low values of [Fe/H], also possibly related to the interaction-driven starbursts triggered by that event.

The overall shape of this distribution also appears to differ between each disc, although the more similarly barred IsoB and TideB are again most homogeneous, compared to TideNC which is again more disparate in features. The most obvious deviation in shape is the elongated tail with higher fractions of [Fe/H] ($\ge 0.3$) for a given [O/Fe] abundance in IsoB and TideB that is conspicuously absent from TideNC. This feature appears to be comprised of predominantly younger stars with formation radii almost exclusively in the inner $\sim 1$\,kpc. The intersection of this component with the main diagonal feature occurs at stellar ages almost exactly consistent with bar formation (IsoB: $\sim 400$\,Myr; TideB: $\sim 200$\,Myr). This implies that the conditions necessary to generate such a feature should be related to the formation of the bar in these two discs but not related to the formation of the bar in TideNC. Based on the formation radii in particular, it could perhaps be assumed that this be related to the small, circular nucleus which is present in IsoB and TideB but not in TideNC in the face-on metallicity projection (see Figure \ref{f:ZmetalPrj}). However, it may also be that the higher star formation in the earlier epochs of IsoB and TideB (with earlier bar formation) simply start to produce more Type Ia Supernovae (SNe Ia) and that TideNC, with peak star formation not occurring until later epochs, does not have enough time to produce these SNe Ia in the relatively short 1\,Gyr simulation period. 

Other simulation studies of both dwarf and Milky Way-like disc galaxies have produced similar features in the abundance ratios \citep[e.g.][]{Bellardini2021, Patel2022, Parul2023}. \citet{Parul2023} find a spreading of the abundance ratios into two tails with distinctly higher and lower values of [O/Fe] for a given [Fe/H] is relatively common for older populations of stars with formation times associated with bursty-type star formation in their sample of Milky-way mass galaxies from the \textsc{FIRE-2} simulations. These simulations are similarly Lagrangian, using the \textsc{gizmo} hydrodynamical code with similar internal physics \citep[see][]{Hopkins2015, Hopkins2018, Parul2023}. They argue that the existence of an upper arm in this regime should be related to the yield tracks of feedback from core-collapse supernovae and winds from AGB stars, as any intense period of star formation must drive rapid enrichment of the ISM leading to the higher fraction of alpha elements traced by [O/Fe]. Comparatively, the lower arm, which is more dominant in Figure \ref{f:OxFe}, should indeed be related to the feedback from SNe Ia. If star formation is considerably bursty, stars must form in small dense clumps rapidly, then stellar feedback will generate strong outflows, quickly deplete available gas and consequently halt star formation. The lower metallicity is a result of the gas from these outflows mixing with halo gas before returning to the disc to once more commence forming stars which will be enriched by these SNe Ia. It is likely that the two distinct arms from separately due to the relative delay between the start of star formation and the first supernovae explosions, as well as a possibly turbulent ISM, which is characteristic of discs sporting bursty type star formation, preventing effective chemical mixing on the timescales of each burst \citep{Hayward2017, Parul2023}. That is, the star formation in this period may be occurring in bursts with times shorter than the local mixing time required for metals produced by either core-collapse supernovae or Type Ia supernovae to sufficiently combine within the surrounding gas. 

However, there is no halo gas included in these single disc type simulations with which the outflows can mix. Additionally,
from the star formation histories of the three discs (IsoB, TideB \& TideNC) neither of these discs appear to be any more, or less, bursty on average than the other. TideNC, for example, is missing this extra feature in Figure \ref{f:OxFe} and yet, does not appear significantly different in the smoothness of the star formation history (see Appendix \ref{a:sf_history}). However, there is at least one significantly sharp increase (or burst) in the SFR which occurs upon bar formation. This burst is indeed much more significant in IsoB and TideB than in TideNC. We suggest that these two arms should trace the chemical tracks in the inner nucleus region of each disc. The initial bar formation, with associated rapid star forming period, should first drive the chemical enrichment toward higher values of [Fe/H] and [O/Fe]. Then, the subsequent infall of lower metallicity gas post-bar formation can suppress the higher star formation occurring within the bar and will flow into the NSD from the bar-edges. This brings lower values for [Fe/H] and [O/Fe]. Finally, the SNe Ia will start generating a path with higher [Fe/H] and lower [O/Fe]. For SNe Ia to lower values of [O/Fe], this decrease in SFR is required. Conversely, a constant SFR can maintain the higher values of [O/Fe] and is likely related to the difference in shape of this abundance relation in TideNC. This chemical evolution of the NSD and whether it is truly varying based on bar formation mechanism is certainly worth further analysis, especially in simulations with more realistic initialisation for disc metallicity. 

\begin{figure}
	\includegraphics[width=0.98\columnwidth]{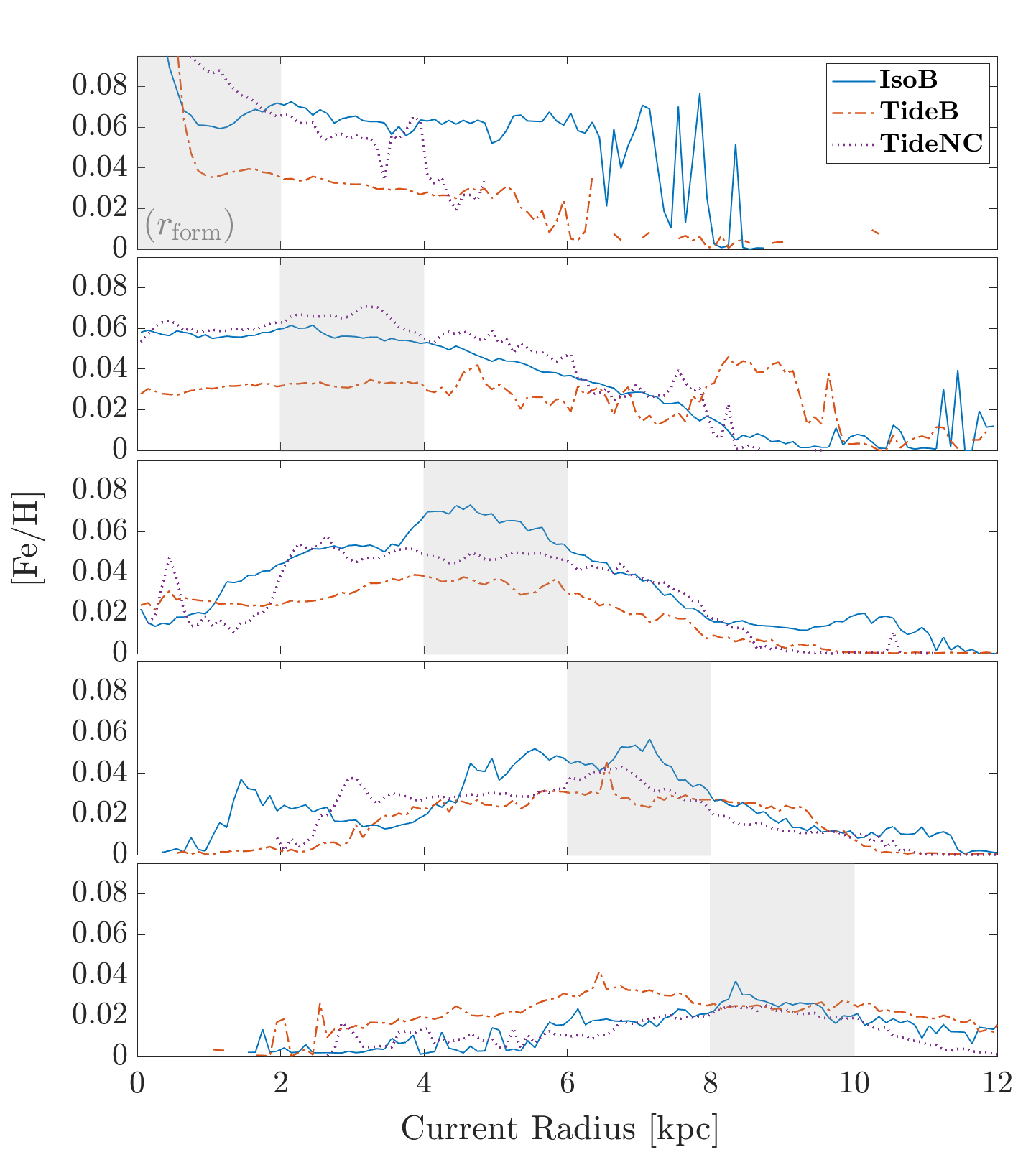}
  \caption{Metallicity distribution ([Fe/H]) with radial position ($r$) at 1\,Gyr for each: IsoB (solid blue line), TideB (dot-dashed orange line) and TideNC (dotted purple line). The five panels are each 2\,kpc bins of star formation radius ($r_{\rm{form}}$) for stars formed in the simulation. The radii which correspond to the formation radius of a given each panel is shaded in grey. }
  \label{f:OxRibinR}
\end{figure}

In an attempt to combine measurements of metallicity with radial migration, Figure \ref{f:OxRibinR} is a representation of the radial dependence for [Fe/H] tracing both the current (1\,Gyr) radial distribution of stars in each disc but also any relation with formation radii. This figure is divided into five panels corresponding to bins of formation radii spanning $r_{\rm form}$ between [0-2], [2-4], [4-6], [6-8] and [8-10]\,kpc respectively (shaded grey, increasing outwards from upper to lower panels). There is no consideration for possible dependence on stellar age; contributions from all stars formed by 1\,Gyr are included. For this figure, any contribution at current radii ($r$) outside of the grey shaded region can be attributed to stellar populations which have moved significantly from the original location and spread different metallicities through the disc. 

Stars formed at different radii appear to contribute to the metallicity in regions of the current disc differently, with also distinguishably different profiles based on the bar formation mechanism in each disc. The radial dependence of the abundance ratio [Fe/H] for stars formed in the two isolated discs (IsoB \& TideNC) appear mostly similar in shape and scale of the distribution. This is particularly evident for stars formed between 2-4 and 8-10\,kpc; that is, at the bar edges and in the outer regions of the disc. The overall magnitude of the [Fe/H] values in the distribution of each component is also generally higher in isolated discs and lower in the tidally-driven disc, except for those stars formed in the outer-regions of the interacting TideB ($r_{\rm{form}} \gtrsim 6$\,kpc). This can perhaps be expected due to stronger effects of tidal forces on the outer-edges from the interaction causing larger changes in position, and thus more significant population spreading for stars formed in these regions. Stars formed in the 8-10\,kpc bin of TideB appear to contribute more to the [Fe/H] fraction at inner radii than to the relative abundances at the formation radii. There is, however, a notable exception to this trend that cannot be immediately explained by lingering interaction effects. There is a significant component of the stars formed between 2-4\,kpc in the TideB disc which have migrated to the far outer reaches of the galactic disc ($r \ge 8$\,kpc). This 2-4\,kpc bin is located precisely where the previously identified separation in the most likely value of $\Delta r$, $\Delta L_z$ and $\Delta z$ for a given radius ($r_i$) is located. It is also the location of the bar-edges and where the largest changes to orbital eccentricity are observed. Additionally, the strange component with large downward vertical motion out of the disc-plane is also predominantly composed of stars formed at these radii. However, as the time of formation is not taken into account in this figure, it is perhaps not possible to directly link these features currently. The specific stars associated with these attributes should be independently extracted and specifically traced through the evolution of the tidally affected disc in future work. 

Other than this odd feature in the 2-4\,kpc bin of TideB, the shape of the distribution in each of the three discs can be considered relatively similar for $r \le 8$\,kpc. The primary affect of the different bar origins should, therefore, be to raise or lower the overall magnitude of the abundance ratio across the disc. It could be assumed then, that a barred-galaxy with an interaction driven formation mechanism will be characterised by a greater difference in the ratio of metallicity between the central/bar and disc regions; with likely evidence of a population of stars in the outer-disc contributing metallicities more similar to the bar/bar-edge conditions and/or on more elongated or eccentric orbits. 

\section{Conclusions}
\label{s:conc}
We have probed three simulations of barred-disc galaxies, with two matching isolated evolutionary histories and one tidally driven disc. Similar bar morphology is produced in each one isolated and one tidal scenario, while the second isolated disc, evolved from the tidal initial condition sans companion, is visibly dissimilar. This work aimed to determine whether the mechanisms driving bar formation could affect the stellar dynamics and migration within these galaxies in significant and discernible ways. If so, it should be possible to use such traits in the stellar populations and kinematics from observations to identify the evolutionary histories of nearby barred-galaxies. 

With these three examples, it is sufficient to demonstrate that bar formation clearly impacts radial migration and vertical action. Similar bars (IsoB \& TideB) drive similar changes to the radial positions and angular momentum in the disc-plane. This is particularly evident in the change to vertical positions ($\Delta z$), where the formation of a bar in face-on morphology also corresponds to large vertical changes in stellar position and an X-like feature is formed in the edge-on perspective which is not evident in the TideNC disc. The traits and magnitude of changes for this TideNC disc are generally smaller and often do not exhibit the same properties as the other two discs. However, all three discs indicate that the disc resonances, particularly co-rotation, are significant pre-bar formation and continue to be associated with large migration features post-bar formation, along with the radial location of the bar-edges. The inner-Lindblad resonance is most significant for changes in orbital eccentricity, particularly in the morphologically dissimilar TideNC, with all discs showing large positive changes in eccentricity preceding bar formation within this resonance position but the feature continuing into the barred periods of TideNC. The impact of the tidal interaction is seen most clearly in population dependence of migration with changes pre- and post-closest approach of the companion and also on the outer-edges of the disc, as could be expected. However, the overall magnitude of change in all features does also appear largest in TideB for most epochs of structure formation. Additionally, there does appear to be a small population of stars formed at inner radii ($\sim 2-4$\,kpc; about where the bar will eventually form) in the first 200\,Myr which is significantly affected by the companion passage. This population appears to contribute large radial migration outward and large downward migration to well-below the disc-plane.

Tracing the different effects of these changes through the evolution of disc metallicity indicates that similar bar morphology should be the most significant in the evolution of metallicity features within the central region. The formation of a clear, high metallicity circular nucleus forms for only IsoB and TideB within the 1\,Gyr period, while TideNC remains a long, thin bar of highest metallicity in the central region. Additionally, the chemical enrichment in TideNC has only one component whereas, IsoB and TideB are dominated by a two-arm feature. It is suggested that this should be related to the timing and strength of the initial rapid star forming period associated with bar formation in each of these discs and that, as TideNC forms a thinner bar at much later epochs, the SNe Ia have not yet sufficiently affected the abundance patterns in this way. However, the full disc metallicity distribution most notably differs based on bar origins. The tidally driven evolutionary scenario in TideB leads to a much larger gradient in metallicity across the disc with almost 25\% greater disparity between the central and outer-disc metallicities, when compared to the two isolated discs (IsoB \& TideNC) that develop very similar ratios despite the larger difference in SFR for these discs. The population of stars in TideB which are formed between 2-4\,kpc and migrate large distances are also notable, contributing to the overall radial dependence of the disc metallicity with higher values of [Fe/H] in the outer-edges of the disc ($r_{\rm{form}}\ge 8$\,kpc) when compared to stars with similar formation radii in each of the isolated discs and even stars formed at other radii within the tidally affected disc.

It has been possible to identify traits in the radial and vertical motion of stellar populations which may distinguish the evolutionary history of galaxies, by way of bar origin. In theory, it should be possible to use these traits in measurements of stellar kinematics and metallicity from observations of real galaxies to distinguish between bars formed in isolation and those triggered by tidal forces. However, we note that the tidal interaction in this case is specifically contrived to be independently distinguishable, while the interaction histories of real galaxies are often complex and it is currently unknown to what extent this may blur or, conversely, strengthen the effects identified herein. Additionally, for a similar reason, the longevity of these features is also beyond the scope of the present study and it is possible that these distinctions may only be measurable in younger bars, observed at earlier stages of their evolution. Nevertheless, we look forward to further research in this area by way of both real and mock observational measurements.

The results presented herein also demonstrate the significance of a given bar morphology on the disc stellar population, therefore, this too should not be disregarded in future studies. The primary morphological difference between the bars in this work is considered to be associated with the lack of bar-buckling in TideNC and, while it has been speculated that buckling may eventually occur in this disc, the features identified as arising from such a difference remains nevertheless significant in the \textquoteleft present\textquoteright \ epoch of observation. As bar formation is not instantaneous, it may be such that the differences in observed bar morphologies could be attributed to differing initial disc conditions, differing interaction histories, or simply differing stages of bar evolution. The third possibility related to bar evolution is not within the scope of this study, except only to indicate that, regardless of whether buckling occurs at a later epoch of TideNC, the differences attributable to morphology in the present epoch should be observable and thus, physically significant.

An increased number of discs with different observed bar morphologies and interaction histories, evolved over a variety of timescales should also be important for constraining the universal consistency and overall relevance of these identified features. The specific evolution in the central regions of barred-galaxies, particularly the chemical properties of the NSD, as well as the bar-located stellar kinematics are raised as significant areas for further research. At present, these results provide both evidence and impetus for using the observable properties of stellar positions and kinematics, ages and metallicities as tracers for the specific origins of a galactic bar and consequently, the opportunity for determining the evolutionary histories of bars in observations of real galaxies.

\section*{Acknowledgements}
We thank the anonymous referee whose comments helped improved the quality of this manuscript. EJI acknowledges the support of Japanese Government MEXT Scholarship for Foreign Students. TO acknowledges the support of MEXT/JSPS KAKENHI Grant 18H04333, 19H01931, and 20H05861. DK acknowledges the support of MWGaiaDN, a Horizon Europe Marie Sk\l{}odowska-Curie Actions Doctoral Network funded under grant agreement no. 101072454 and also funded by UK Research and Innovation (EP/X031756/1), and the UK's Science \& Technology Facilities Council (STFC grant ST/W001136/1). Images and partial analysis were made using the \textsc{pynbody} \textsc{python} package \citep[https://github.com/pynbody/pynbody , ][]{Pontzen2012} and the MathWorks \textsc{matlab} programming and numeric computing platform. Observational data was accessed from the PHANGS-HST survey \citep{Lee2021}. Numerical computations were carried out on the Cray XC50 at Center for Computational Astrophysics, National Astronomical Observatory of Japan.

\section*{Data Availability}
The data underlying this article can be shared upon reasonable request to the corresponding author.

\bibliographystyle{mnras}
\bibliography{AA_main}

\newpage
\appendix

\section{Star Formation History}
\label{a:sf_history}
\begin{figure}
	\includegraphics[width=\columnwidth]{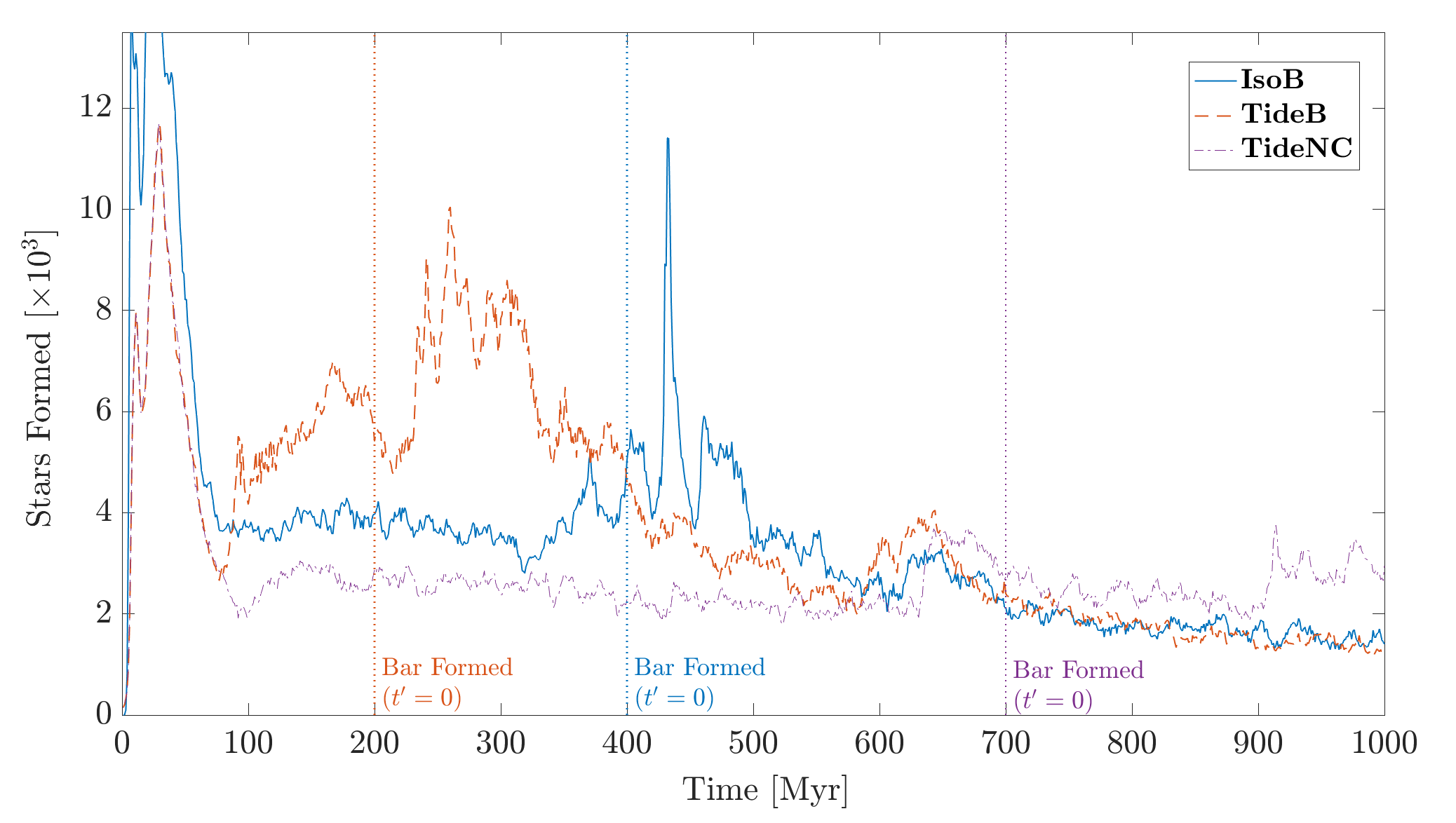}
  \caption{Star formation history for each: IsoB (solid blue line), TideB (orange dashed line) and TideNC (purple dot-dashed line). The axis shows simulation time up to 1\,Gyr with a vertical line to denotes the time of bar formation (\textsc{NB:} the large peak at $t < 100$\,Myr is an artefact of the simulation).}
  \label{f:sf_history}
\end{figure}

To determine the bursty nature of star formation in these discs, the star formation history over the 1\,Gyr simulation time is presented in Figure \ref{f:sf_history}

\bsp	
\label{lastpage}
\end{document}